\documentclass[aps,prd,showpacs,twocolumn,groupedaddress]{revtex4}
\usepackage{graphicx}
\usepackage{amsfonts}
\usepackage{amssymb}
\usepackage{amsmath}
\usepackage{bm}

\usepackage{ulem}
\usepackage{color}

\begin{document}

\title{Relation between Confinement and Chiral Symmetry Breaking
\\  in Temporally Odd-number Lattice QCD}

\author{Takahiro~M.~Doi}
  \email{doi@ruby.scphys.kyoto-u.ac.jp}
  \affiliation{Department of Physics, Graduate School of Science,
  Kyoto University, \\
  Kitashirakawa-oiwake, Sakyo, Kyoto 606-8502, Japan}
\author{Hideo~Suganuma}
  \email{suganuma@ruby.scphys.kyoto-u.ac.jp}
  \affiliation{Department of Physics, Graduate School of Science,
  Kyoto University, \\
  Kitashirakawa-oiwake, Sakyo, Kyoto 606-8502, Japan}
\author{Takumi~Iritani}
  \email{iritani@post.kek.jp}
  \affiliation{High Energy Accelerator Research Organization (KEK), \\
  Tsukuba, Ibaraki 305-0801, Japan}

\date{\today}
\begin{abstract}
In the lattice QCD formalism, 
we investigate the relation 
between confinement and chiral symmetry breaking. 
A gauge-invariant analytical relation connecting the Polyakov loop and the Dirac modes is derived 
on a temporally odd-number lattice, where the temporal lattice size is odd, 
with the normal (nontwisted) periodic boundary condition 
for link-variables. 
This analytical relation indicates that low-lying Dirac modes have little contribution to the Polyakov loop, 
and it is numerically confirmed at the quenched level in both confinement and deconfinement phases. 
This fact indicates no direct one-to-one correspondence 
between confinement and chiral symmetry breaking in QCD. 
Using the relation, we also investigate the contribution from each Dirac mode to the Polyakov loop. 
In the confinement phase, we find a new  ``positive/negative symmetry" 
of the Dirac-mode matrix element of the link-variable operator, 
and this symmetry leads to the zero value of the Polyakov loop. 
In the deconfinement phase, there is no such symmetry and the Polyakov loop is nonzero. 
Also, we develop a new method for spin-diagonalizing the Dirac operator 
on the temporally odd-number lattice 
modifying the Kogut-Susskind formalism. 

\end{abstract}
\maketitle

\def\slash#1{\not\!#1}
\def\slashb#1{\not\!\!#1}
\def\slashbb#1{\not\!\!\!#1}

\section{Introduction}
Color confinement and chiral symmetry breaking are very important phenomena 
in nuclear and elementary particle physics and 
have been investigated as interesting non-perturbative phenomena 
in low-energy QCD in many analytical and numerical studies \cite{NJL, KS, Rothe, Greensite}. 
However, their properties are not sufficiently understood directly from QCD. 
The Polyakov loop is an order parameter for quark confinement \cite{Rothe}. 
At the quenched level, the Polyakov loop is the exact order parameter for quark confinement, 
and its expectation value is zero in the confinement phase 
and nonzero in the deconfinement phase. 
As for chiral symmetry, the order parameter of chiral symmetry breaking is chiral condensate, and low-lying Dirac modes are essential for chiral symmetry breaking in QCD, 
for example, according to the Banks-Casher relation \cite{BanksCasher}. 

Not only the properties of confinement and chiral symmetry breaking in QCD 
but also their relation is an interesting challenging subject 
\cite{SST95, Miyamura, Woloshyn, Hatta, Langfeld, Gattringer, Karsch, YAoki, GIS, BG08, LS11}. 
From some studies, it is suggested that confinement and chiral symmetry breaking are strongly correlated. 
In finite temperature lattice QCD calculation, 
some studies tell that 
the transition temperatures of deconfinement phase transition and chiral restoration 
are almost the same \cite{Karsch}. 
Also, by removing QCD monopoles in the maximally Abelian gauge, both confinement and chiral symmetry breaking 
are simultaneously lost in lattice QCD \cite{Miyamura, Woloshyn}. 
However, there is an opposite study that 
the transition temperatures of deconfinement phase transition 
and chiral restoration are not the same \cite{YAoki}. 

In recent lattice-QCD numerical studies, 
it is suggested that the properties of confinement are not changed 
by removing low-lying Dirac modes from the QCD vacuum \cite{GIS}. 
Since low-lying Dirac modes are essential for chiral symmetry breaking, 
this calculation indicates that 
there is no one-to-one correspondence between confinement and chiral symmetry breaking in QCD. 

To investigate the relation between confinement and chiral symmetry breaking, 
the analytical relation between the Polyakov loop and Dirac modes is very useful. 
For example, the Polyakov loop is expressed in terms of Dirac eigenvalues 
under the twisted boundary condition for link-variables \cite{Gattringer}. 
However, the (anti)periodic boundary condition is physically important 
for the imaginary-time formalism at finite temperature. 
Recently, we derived a relation between the Polyakov loop and Dirac modes 
on a temporally odd-number lattice, where the temporal lattice size is odd, 
with the normal nontwisted periodic boundary condition for link variables \cite{SDI, DSI}. 

In this study, we analytically and numerically investigate 
the relation between confinement and chiral symmetry breaking. 
In Sec. II, we derive an analytical relation 
connecting the Polyakov loop and Dirac modes on the temporally odd-number lattice. 
In Sec. III, we develop a new method for spin-diagonalizing the Dirac operator applicable to the temporally odd-number lattice modifying the Kogut-Susskind (KS) formalism \cite{KS}. 
In Sec. IV, for more detailed analysis, we perform the numerical analysis based on the relation. 
Section V is summary and discussion. 

\section{the relation between the Polyakov loop and Dirac modes on the temporally odd-number lattice}
In this section, 
we derive the relation between the Polyakov loop and Dirac modes on the temporally odd-number lattice 
with the normal (nontwisted) periodic boundary condition for link-variables 
in both temporal and spatial directions \cite{SDI, DSI}. 
\subsection{Operator formalism and Dirac mode in lattice QCD}
As the preparation, we review operator formalism and Dirac modes in the SU($N_{\rm c}$) lattice QCD. 
We use a standard square lattice with spacing $a$, and the notation of 
sites $s=(s_1, s_2, s_3, s_4) \ (s_\mu=1,2,\cdots,N_\mu)$, 
and link-variables $U_\mu(s)={\rm e}^{iagA_\mu(s)}$ 
with gauge fields $A_\mu(s) \in su(N_c)$ and gauge coupling $g$.
In this paper, we define all the $\gamma$-matrices to be hermite as $\gamma^\dagger_\mu=\gamma_\mu$.

We define the link-variable operator $\hat{U}_{\pm\mu}$ 
by the matrix element,
\begin{align}
\langle 
s | \hat{U}_{\pm\mu} |s' \rangle=U_{\pm\mu}(s)\delta_{s\pm\hat{\mu},s'}, \label{LinkOp}
\end{align}
where $\hat{\mu}$ is the unit vector in direction $\mu$ in the lattice unit.
Using the link-variable operator, 
the Polyakov loop $L_P$ is expressed as 
\begin{align}
L_P
&=\frac{1}{3V}{\rm Tr}_c \{\hat U_4^{N_4}\} \label{PolyakovOp} \nonumber \\
&=\frac{1}{3V} \sum_s {\rm tr}_c
\{\prod_{i=0}^{N_4-1}U_4(s+i\hat{4})\},
\end{align}
with the 4D lattice volume $V=N_1N_2N_3N_4$. 
Here, ``Tr$_c$'' denotes the functional trace of 
${\rm Tr}_c \equiv \sum_s {\rm tr}_c$
with the trace ${\rm tr}_c$ over color index. 

Also, using the link-variable operator, 
covariant derivative operator $\hat{D}_\mu$ 
on the lattice is expressed as
\begin{align}
\hat{D}_\mu= \frac{1}{2a}(\hat{U}_{\mu}-\hat{U}_{-\mu}). \label{CovariantOp}
\end{align}
Thus, in the lattice QCD, the Dirac operator $\hat{\slashb{D}}$ is expressed as 
\begin{align}
\hat{\slashb{D}}=\gamma_\mu \hat{D}_\mu= 
\frac{1}{2a}\sum_{\mu=1}^{4}\gamma_\mu (\hat{U}_{\mu}-\hat{U}_{-\mu}), \label{DiracOp}
\end{align}
and its matrix element is explicitly expressed as 
\begin{align}
 \slashb{D}_{s,s'} 
      = \frac{1}{2a} \sum_{\mu=1}^4 \gamma_\mu 
\left[ U_\mu(s) \delta_{s+\hat{\mu},s'}
        - U_{-\mu}(s) \delta_{s-\hat{\mu},s'} \right], \label{DiracOpExp}
\end{align}
with $U_{-\mu}(s)\equiv U^\dagger_\mu(s-\hat\mu)$.
Since the Dirac operator is anti-hermite in this definition of $\gamma_\mu$, 
the Dirac eigenvalue equation is expressed as 
\begin{eqnarray}
\hat{\slashb{D}}|n\rangle =i\lambda_n|n \rangle
\end{eqnarray}
with the Dirac eigenvalue $i\lambda_n$ ($\lambda_n \in {\bf R}$) 
and the Dirac eigenstate $|n \rangle$. 
These Dirac eigenstates have the completeness of $\sum_n|n \rangle\langle n|=1$. 
According to $\{\hat{\slashb{D}},\gamma_5\}=0$,
the chiral partner $\gamma_5|n\rangle$ is also 
an eigenstate with the eigenvalue $-i\lambda_n$.
Using the Dirac eigenfunction $\psi_n(s)\equiv\langle s|n \rangle $, 
the explicit form for the Dirac eigenvalue equation is written by 
\begin{eqnarray}
 \frac{1}{2a}& \sum_{\mu=1}^4 \gamma_\mu
[U_\mu(s)\psi_n(s+\hat \mu)-U_{-\mu}(s)\psi_n(s-\hat \mu)] \ \ \ \ \ & \nonumber \\
& \ \ \ \ \ \ \ \ \ \ \ \ \ \ \ \ \ \ \ \ \ \ \ \ \ \ \ \ \ \ \ \ \ \ \ \ \ \ \ \ \ \ \ =i\lambda_n \psi_n(s). 
\label{DiracEigenExp}
\end{eqnarray}
The Dirac eigenfunction $\psi_n(s)$ can be numerically obtained in lattice QCD, 
besides a phase factor. 
By the gauge transformation of $U_\mu(s)\rightarrow V(s)U_\mu(s)V^\dagger(s+\hat{\mu})$,
$\psi_n(s)$ is gauge-transformed as
\begin{eqnarray}
\psi_n(s)\rightarrow V(s)\psi_n(s), \label{GaugeTrans}
\end{eqnarray}
which is the same as that of the quark field, although, to be strict, 
there can appear an irrelevant $n$-dependent 
global phase factor $\mathrm{e}^{i\varphi_n[V]}$, 
according to arbitrariness of the phase in the basis $|n\rangle$ \cite{GIS}.

The Dirac-mode matrix element of the link-variable operator $\hat{U}_\mu$ can be expressed with $\psi_n(s)$: 
\begin{align}
\langle m|\hat{U}_\mu| n \rangle  
&=
\sum_s \langle m|s \rangle \langle s|\hat{U}_\mu| s+\hat{\mu} \rangle \langle s+\hat{\mu}|n \rangle \nonumber \\
&=
\sum_s \psi_m^\dagger(s)U_\mu(s)\psi_n(s+\hat{\mu}). \label{MatEleExp}
\end{align}
Note that the matrix element is gauge invariant, apart from an irrelevant phase factor. 
Actually, using the gauge transformation Eq.(\ref{GaugeTrans}), we find the gauge transformation of
the matrix element as \cite{GIS}
\begin{align}
&\langle m|\hat{U}_\mu| n \rangle 
=
\sum_s \psi_m^\dagger(s)U_\mu(s)\psi_n(s+\hat{\mu}) \nonumber \\ 
&\rightarrow
\sum_s \psi_m^\dagger(s)V^\dagger(s)\cdot V(s)U_\mu(s)V^\dagger(s+\hat{\mu}) \nonumber \\ 
&\cdot V(s+\hat{\mu})\psi_n(s+\hat{\mu}) \nonumber \\ 
&=
\sum_s \psi_m^\dagger(s)U_\mu(s)\psi_n(s+\hat{\mu})
=\langle m|\hat{U}_\mu| n \rangle. 
\end{align}
To be strict, there appears an $n$-dependent global phase factor, corresponding to the arbitrariness of the phase
in the basis $|n\rangle$. However, this phase factor cancels as 
$\mathrm{e}^{i\varphi_n[V]}\mathrm{e}^{-i\varphi_n[V]}=1$ 
between $|n \rangle$ and $\langle n|$, and does not appear
for physical quantities such as the Wilson loop and the Polyakov loop \cite{GIS}.

Note also that a functional trace of a product of the link-variable operators 
corresponding to the non-closed path is exactly zero 
because of the definition of 
the link-variable operator Eq.(\ref{LinkOp}): 
\begin{align}
&{\rm Tr}_c(\hat{U}_{\mu_1}\hat{U}_{\mu_2}\cdots\hat{U}_{\mu_{N}})
=
{\rm tr}_c\sum_s\langle s|\hat{U}_{\mu_1}\hat{U}_{\mu_2}\cdots\hat{U}_{\mu_{N}}|s\rangle \nonumber \\
=&{\rm tr}_c\sum_s U_{\mu_1}(s)U_{\mu_2}(s+\hat{\mu}_1)\cdots U_{\mu_{N}}(s+\sum_{k=1}^{N-1}\hat{\mu}_k)
\langle s+\sum_{k=1}^{N}\hat{\mu}_k |s\rangle \nonumber \\
=&0
\label{nonclosed}
\end{align}
with $\sum_{k=1}^{N}\hat{\mu}_k\neq 0$ 
for the non-closed path and the length of the path $N$. 
This is easily understood from Elitzur's theorem \cite{Elitzur} that 
the vacuum expectation values of gauge-variant operators are zero. 

Dirac modes are strongly related to the chiral condensate according to 
the Banks-Casher relation \cite{BanksCasher}: 
\begin{align}
\langle\bar{q}q\rangle=-\lim_{m\to0}\lim_{V\to\infty}\pi\langle\rho(0)\rangle, \label{BCrel} 
\end{align}
where the Dirac eigenvalue density $\rho(\lambda)$ is defined by 
\begin{align}
\rho(\lambda)\equiv \frac{1}{V_{\rm phys}}\sum_n\langle\delta(\lambda-\lambda_n)\rangle
\label{BCrel} 
\end{align}
with the space-time volume $V_{\rm phys}$. 
From Eq.(\ref{BCrel}), the chiral condensate is proportional to the Dirac zero-eigenvalue density. 
Since the chiral condensate is the order parameter 
of chiral symmetry breaking, low-lying Dirac modes are essential 
for chiral symmetry breaking. 
In general, instead of $\slashb{D}$, one can consider any (anti)hermitian operator, e.g.,
$D^2 =D_\mu D_\mu$, and the expansion in terms of its eigen-modes \cite{BI05}. 
To investigate chiral symmetry breaking, however, it is appropriate to consider $\slashb{D}$ 
and the expansion by its eigenmodes. 

Note here that, although the Polyakov loop is defined by gauge fields alone, 
there can be some relation to the Dirac modes, as will be shown later. 
For, the Dirac modes are strongly affected by the gauge fields.
A similar example is instantons. 
The instantons are defined by gauge fields alone; 
however they have a close connection to the axial U(1) anomaly, 
which relates to a fermionic symmetry.
In fact, even though the Polyakov loop is defined by gauge fields alone, 
it has a physical meaning to consider the relation to 
some fermionic modes in QCD.

The role of the low-lying Dirac modes has been studied 
in the context of chiral symmetry breaking in QCD. 
In particular, the removal of low-lying Dirac modes has been recently 
investigated to realize 
the world of ``unbreaking chiral-symmetry" \cite{LS11,GIS}. 
For example, propagators and masses of hadrons are investigated 
after the removal of low-lying Dirac modes, 
and parity-doubling ``hadrons" can be actually observed as bound states 
in the chiral unbroken world \cite{LS11}. 
Also, after the removal of low-lying Dirac modes from the QCD vacuum, 
the confinement properties such as the string tension are found to be 
almost kept, while the chiral condensate is largely decreased \cite{GIS}.

\subsection{The relation between Polyakov loop and Dirac modes on the temporally odd-number lattice}
We consider the temporally odd-number lattice, where the temporal lattice size $N_4$ is odd, with 
the normal (nontwisted) periodic boundary condition 
for link-variables in both temporal and spatial directions. 
The spatial lattice size $N_{1 \sim 3} (> N_4)$ is taken to be even. 

\begin{figure}[h]
\begin{center}
\includegraphics[scale=0.4]{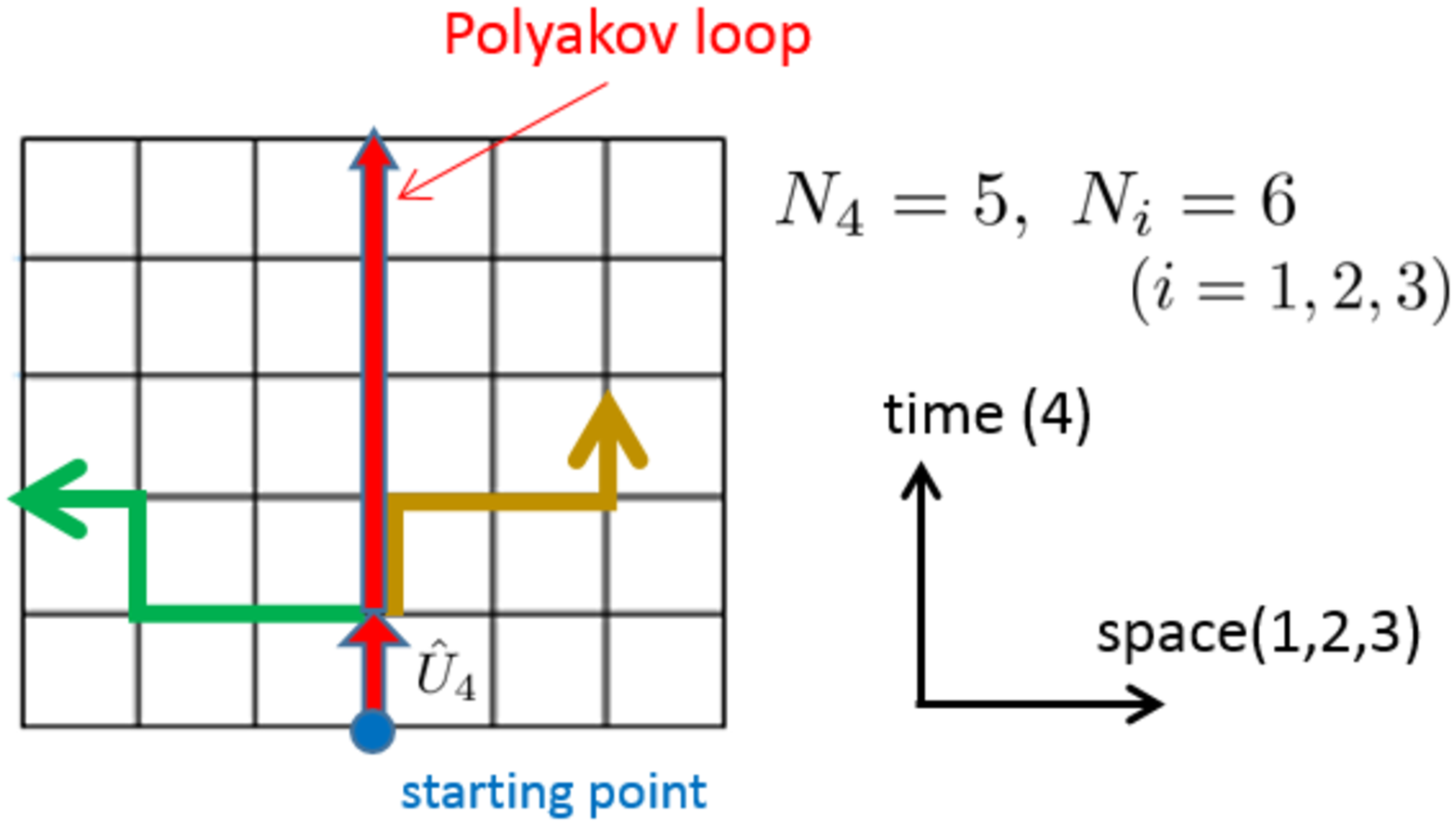}
\caption{
An example of temporally odd-number lattice. 
This is $N_4=5, \ N_i=6 \ (i=1,2,3)$ case. 
Each line corresponds to each term in $\hat{U}_4\hat{\slashb{D}}^{N_4-1}$ in Eq.(\ref{I}). 
On square lattice, one cannot make any closed loops using products of odd-number link-variable operators. 
}
\label{OddLattice}
\end{center}
\end{figure}

First, as a key quantity, we introduce
\begin{eqnarray}
I\equiv {\rm Tr}_{c,\gamma} (\hat{U}_4\hat{\slashb{D}}^{N_4-1}) \label{I}
\end{eqnarray}
with the functional trace 
${\rm Tr}_{c,\gamma}\equiv \sum_s {\rm tr}_c 
{\rm tr}_\gamma$ including also 
the trace ${\rm tr}_\gamma$ over spinor index. 
From Eq.(\ref{DiracOp}), $\hat{U}_4\hat{\slashb{D}}^{N_4-1}$ is expressed as 
a sum of products of $N_4$ link-variable operators. 
In Fig. \ref{OddLattice}, an example of the temporally odd-number lattice is 
shown and 
each line corresponds to each term in $\hat{U}_4\hat{\slashb{D}}^{N_4-1}$ in Eq.(\ref{I}). 
Here, note that 
one cannot make any closed loops using products of odd-number link-variable operators on a square lattice. 
Since now $N_4$ is odd and we consider the square lattice, 
$\hat{U}_4\hat{\slashb{D}}^{N_4-1}$ does not have any operators corresponding to closed paths 
except for the term proportional to $\hat{U}_4^{N_4}$ 
which corresponds to a closed path and is gauge invariant 
because of the periodic boundary condition for time direction, 
which is proportional to the Polyakov loop. 
Therefore using Eqs.(\ref{PolyakovOp}), (\ref{CovariantOp}) and (\ref{nonclosed}), 
we obtain 
\begin{align}
I
&={\rm Tr}_{c,\gamma} (\hat{U}_4 \hat{\slashb{D}}^{N_4-1}) \nonumber \\
&={\rm Tr}_{c,\gamma} \{\hat{U}_4 (\gamma_4 \hat{D}_4)^{N_4-1}\} \nonumber \\
&=4 {\rm Tr}_{c} (\hat{U}_4 \hat{D}_4^{N_4-1}) \nonumber \\
&=\frac{4}{(2a)^{N_4-1}}{\rm Tr}_{c} \{\hat{U}_4 (\hat{U}_4-\hat{U}_{-4})^{N_4-1}\} \nonumber \\
&=\frac{4}{(2a)^{N_4-1}}{\rm Tr}_{c} \{ \hat{U}_4^{N_4} \} \nonumber \\
&=\frac{12V}{(2a)^{N_4-1}}L_P. \label{I1}
\end{align}

On the other hand, 
taking Dirac modes as the basis for the functional trace in Eq.(\ref{I}), 
we find 
\begin{align}
I
&=\sum_n\langle n|\hat{U}_4\slashb{\hat{D}}^{N_4-1}|n\rangle \nonumber\\
&=i^{N_4-1}\sum_n\lambda_n^{N_4-1}\langle n|\hat{U}_4| n \rangle.  \label{I2}
\end{align}

Combining Eqs.(\ref{I1}) and (\ref{I2}), we obtain a relation between 
the Polyakov loop $L_P$ 
and the Dirac eigenvalues $i\lambda_n$: 
\begin{eqnarray}
L_P=\frac{(2ai)^{N_4-1}}{12V}
\sum_n\lambda_n^{N_4-1}\langle n|\hat{U}_4| n \rangle
.  \label{RelOrig}
\end{eqnarray}
This is a relation directly connecting the Polyakov loop and the Dirac modes, 
i.e., a Dirac spectral representation of the Polyakov loop. 
Since the Polyakov loop is gauge invariant and Dirac modes can be obtained gauge-covariantly, 
this relation is gauge invariant. 
From the relation (\ref{RelOrig}), we can investigate each Dirac mode contribution to the Polyakov loop individually. 

Since the relation (\ref{RelOrig}) is satisfied for each gauge configuration, 
of course, the relation is satisfied for the gauge-configuration average: 
\begin{eqnarray}
\langle L_P\rangle=\frac{(2ai)^{N_4-1}}{12V}
\left\langle
\sum_n\lambda_n^{N_4-1}\langle n|\hat{U}_4| n \rangle
\right\rangle.  \label{RelOrigVEV}
\end{eqnarray}
The outermost bracket $\langle \rangle$ means gauge-configuration average. 

We can discuss the relation between confinement 
and chiral symmetry breaking in QCD from the relation (\ref{RelOrig}). 
Dirac matrix element $\langle n|\hat{U}_4| n \rangle$ is generally nonzero. 
Thus, the contribution from low-lying Dirac modes with $|\lambda_n|\simeq 0$ 
is relatively small in the sum of RHS in Eq.(\ref{RelOrig}),
compared to the other Dirac-mode contribution because of the damping factor $\lambda_n^{N_4-1}$.
In fact, the low-lying Dirac modes have little contribution 
to the Polyakov loop. This is consistent with the previous numerical 
lattice result that confinement properties, such as interquark potential and the Polyakov loop, 
are almost unchanged by removing low-lying Dirac modes from the QCD vacuum \cite{GIS}.
Thus, we conclude from the relation (\ref{RelOrig}) that there is 
no one-to-one correspondence between confinement and chiral symmetry breaking in QCD. 

The relation (\ref{RelOrig}) is valid only on the temporally 
odd-number lattice, but this constraint is not so serious 
because we are interested in continuum QCD and the parity of the lattice size is not important for physics. 
In fact, by a similar manner on Eq.(\ref{RelOrig}), we can also derive 
a relation which connects the Polyakov loop and Dirac modes on the 
even lattice (see Appendix \ref{EvenDerivation}). 

In the derivation of the relation (\ref{RelOrig}), 
we use only the following setup: 
\begin{enumerate}
 \item odd $N_4$
 \item square lattice
 \item temporal periodicity for link-variables
\end{enumerate}
Therefore, the relation (\ref{RelOrig}) is valid in full QCD and in finite temperature and density, 
and furthermore regardless of the phase of the system. 
In other words, the relation (\ref{RelOrig}) holds in confinement and deconfinement phases, 
and in chiral broken and restored phases. 
Of course, the dynamical quark effect appears in the Polyakov loop $L_P$, 
the Dirac eigenvalue distribution $\rho(\lambda)$, 
and the matrix elements $\langle n|\hat{U}_\mu| m \rangle$. 
However, the relation Eq.(\ref{RelOrig}) holds even in the presence of dynamical quarks. 

For quantitative discussion, 
we numerically calculate each term in the relation (\ref{RelOrig}) 
and investigate each Dirac-mode contribution to the Polyakov loop individually. 
Using Dirac eigenfunction $\psi_n(s)$, 
Dirac matrix element $\langle n|\hat{U}_\mu| m \rangle$ is explicitly expressed as Eq.(\ref{MatEleExp}). 
Thus, the relation (\ref{RelOrig}) is expressed as 
\begin{align}
 L_P =\frac{(2ai)^{N_4-1}}{12V}
\sum_n\lambda_n^{N_4-1}
\sum_{s}\psi^\dagger_n(s)
U_4(s) \psi_n(s+\hat{4}). 
\label{RelOrigExp}
\end{align}
Dirac eigenvalues $\lambda_n$ and Dirac eigenfunctions $\psi_n(s)$ in Eq. (\ref{RelOrigExp}) 
can be obtained 
by solving the Dirac eigenequation (\ref{DiracEigenExp}) using link variables in each gauge configuration. 
However, the numerical cost for solving the Dirac eigenequation is very large 
because of the huge dimension of the Dirac operator $(4\times N_{\rm c}\times V)^2$. 
The numerical cost can be partially reduced without approximation 
using the Kogut-Susskind (KS) formalism \cite{KS} discussed in the next section. 

\section{Modified Kogut-Susskind formalism for temporally odd-number lattice}

In our study, we need all the eigenvalues and the eigenmodes of 
the Dirac operator $\slashb{D}$ defined by Eq.(\ref{DiracOp}). 
This can be numerically performed by the diagonalization of $\slashb{D}$. 
Here, to reduce the numerical cost, we use the technique of 
the KS formalism for diagonalizing the Dirac operator $\slashb{D}$.
Note here that this procedure is just a mathematical technique 
to diagonalize $\slashb{D}$, 
and this never means to use a specific fermion like the KS fermion. 
In fact, the diagonalization of $\slashb{D}$ is mathematically 
equivalent to the use of the KS formalism.

The KS formalism is the method for spin-diagonalizing the Dirac operator 
on the lattice. 
However, when the periodic boundary condition is imposed on the lattice, 
the original KS formalism is applicable only to the ``even lattice" 
where all the lattice sizes are even number. 
In this section, modifying the KS formalism, 
we develop the ``modified KS formalism" applicable to the temporally odd-number lattice \cite{DSI}. 

\subsection{Normal Kogut-Susskind formalism for even lattice}
First, we review the original KS formalism and consider the even lattice, 
where all the lattice sizes $N_{1 \sim 4}$ are even number. 
Using a matrix $T(s)$ defined as 
\begin{align}
T(s)\equiv\gamma_1^{s_1}\gamma_2^{s_2}\gamma_3^{s_3}\gamma_4^{s_4}, \label{T}
\end{align}
one can diagonalize all the $\gamma$-matrices $\gamma_\mu \ (\mu=1,2,3,4)$, 
\begin{align}
T^\dagger(s)\gamma_\mu T(s\pm\hat{\mu})=\eta_\mu(s){\bf 1}, \label{TgammaT}
\end{align}
where staggered phase $\eta_\mu(s)$ is defined as 
\begin{align}
\eta_1(s)\equiv 1, \ \ \eta_\mu(s)\equiv (-1)^{s_1+\cdots+s_{\mu-1}} 
\ (\mu \geq 2). \label{eta}
\end{align}
Since the Dirac operator is expressed as $\slashb{D}=\gamma_\mu D_\mu$, 
one can spin-diagonalize the Dirac operator
\begin{align}
&\sum_\mu T^\dagger(s) \gamma_\mu D_\mu T(s+ \hat \mu) \nonumber\\
& \ \ \ \ \ \ \ \ \ \ \ \ \ 
= {\rm diag}(\eta_\mu D_\mu,\eta_\mu D_\mu,\eta_\mu D_\mu,\eta_\mu D_\mu), \label{TDiracT}
\end{align}
where the KS Dirac operator $\eta_\mu D_\mu$ is defined as 
\begin{align}
(\eta_\mu D_\mu)_{ss'}=\frac{1}{2a}\sum_{\mu=1}^{4}\eta_\mu(s)\left[U_\mu(s)\delta_{s+\hat{\mu},s'}-U_{-\mu}(s)\delta_{s-\hat{\mu},s'}\right]. \label{KSDiracOp}
\end{align}
Equation (\ref{TDiracT}) shows fourfold degeneracy of the Dirac eigenvalue 
relating to the spinor structure of the Dirac operator. 
Thus, one can obtain all the eigenvalues of the Dirac operator by solving the KS Dirac eigenvalue equation 
\begin{align}
\eta_\mu D_\mu|n) =i\lambda_n|n ) \label{KSEigenEq}
\end{align}
with the KS Dirac eigenstate $|n)$.  
Since the KS Dirac operator has only indices of sites and colors, 
the numerical cost for solving the KS Dirac eigenvalue equation (\ref{KSEigenEq}) is smaller than 
that for solving the Dirac eigenvalue equation (\ref{DiracEigenExp}). 
Using the KS Dirac eigenfunction $\chi_n(s)\equiv\langle s|n ) $, 
the KS Dirac eigenvalue equation (\ref{KSEigenEq}) is explicitly expressed as 
\begin{align}
&\frac{1}{2a}\sum_{\mu=1}^4 
\eta_\mu(s)[U_\mu(s) \chi_n(s+\hat \mu)-U_{-\mu}(s)
\chi_n(s-\hat \mu)] \ \ \ \ \  \nonumber \\
& \ \ \ \ \ \ \ \ \ \ \ \ \ \ \ \ \ \ \ \ \ \ \ \ \ \ \ \ 
\ \ \ \ \ \ \ \ \ \ \ \ \ \ \ \ \ \  =i\lambda_n\chi_n(s). \label{KSEigenExp}
\end{align}
Also, KS Dirac matrix element $(n|\hat{U}_\mu|m)$ is expressed as 
\begin{align}
(n|\hat{U}_\mu|m)
&=\sum_s (n|s \rangle \langle s|\hat{U}_\mu| s+\hat{\mu} \rangle\langle s+\hat{\mu}|m ) \nonumber\\
&=\sum_s \chi_n(s)^\dagger U_\mu(s) \chi_m(s+\hat{\mu}). 
\end{align}

Because of fourfold degeneracy of the Dirac eigenvalue, 
there are four states whose eigenvalues are the same, 
and we label these states with quantum number $I=1,2,3,4$, 
namely, $|n,I\rangle$ \cite{Rothe}. 
In this notation, the Dirac eigenvalue equation (\ref{DiracEigenExp}) is expressed as 
\begin{align}
\slashb{D}|n,I\rangle =i\lambda_n|n,I \rangle. \label{EigenEqI}
\end{align}
The relation between the Dirac eigenfunction $\psi_n^{I}(s)_\alpha\equiv\langle s, \alpha|n,I\rangle$ 
and the spinless eigenfunction $\chi_n(s)$ is  
\begin{align}
\psi_n^{I}(s)_\alpha=T(s)_{\alpha\beta}C_\beta^{I}\chi_n(s), \label{PsiChiEven1}
\end{align}
where $C$ is defined as 
\begin{align}
C_\alpha^I=\delta_\alpha^I. \label{C}
\end{align}
Substituting Eq. (\ref{C}) for Eq. (\ref{PsiChiEven1}), one can obtain the relation 
\begin{align}
\psi_n^{I}(s)_\alpha=T(s)_{\alpha I}\chi_n(s), \label{PsiChiEven2}
\end{align}
and quantum number $I$ is mixed with spinor indices. 
This is natural result because the quantum number $I$ is caused by 
the fourfold degeneracy of the Dirac eigenvalue 
and is relating the spinor structure of the Dirac operator. 

When one imposes the periodic boundary condition on the lattice, 
the KS formalism is applicable only to the even lattice. 
In fact, the periodic boundary condition of the matrix $T(s)$ is expressed as 
\begin{align}
T(s+N_\mu\hat{\mu})=T(s) \ \ \ \ \ \ \ (\mu=1,2,3,4), 
\end{align}
and this relation is valid only on the even lattice.  
A spatial periodic boundary condition is not necessarily needed physically, 
but a temporal periodic boundary condition is needed for the imaginary-time finite-temperature formalism. 
Therefore, the original KS formalism is not applicable to the temporally odd-number lattice. 

\subsection{Modified Kogut-Susskind formalism for temporally odd-number lattice}
Now, we present the modified KS formalism as the generalization 
applicable to the temporally odd-number lattice, 
where the lattice size for temporal direction $N_4$ is odd number and 
the lattice sizes for spatial direction $N_i \ (i=1,2,3)$ are even number. 

Instead of the matrix $T(s)$, we define a matrix $M(s)$ by 
\begin{align}
M(s)\equiv\gamma_1^{s_1}\gamma_2^{s_2}\gamma_3^{s_3}\gamma_4^{s_1+s_2+s_3}. \label{M}
\end{align}
The matrix $M(s)$ is similar to the matrix $T(s)$, 
but independent of the time component of the site $s_4$. 
Using the matrix $M(s)$, all the $\gamma-$matrices are transformed to be proportional to $\gamma_4$: 
\begin{align}
M^\dagger(s)\gamma_\mu M(s\pm\hat{\mu})=\eta_\mu(s)\gamma_4, \label{MgammaM}
\end{align}
where $\eta_\mu(s)$ is the staggered phase given by Eq. (\ref{eta}). 
In the Dirac representation, $\gamma_4$ is diagonal as 
\begin{align}
 \gamma_4={\rm diag}(1,1,-1,-1) \ \ \ (\rm{Dirac \ representation}),
\label{gamma_4}
\end{align}
and we take the Dirac representation in this paper. 
Thus, one can spin-diagonalize the Dirac operator $\slashbb{D}=\gamma_\mu D_\mu$ 
in the case of the temporally odd-number lattice: 
\begin{align}
&\sum_\mu M^\dagger(s) \gamma_\mu D_\mu M(s+ \hat \mu) \nonumber\\
& \ \ \ \ \ \ \ \ \ \ \ \ \ 
= {\rm diag}(\eta_\mu D_\mu,\eta_\mu D_\mu,-\eta_\mu D_\mu,-\eta_\mu D_\mu), \label{MDiracM}
\end{align}
where $\eta_\mu D_\mu$ is the KS Dirac operator given by Eq. (\ref{KSDiracOp})

As a remarkable feature, 
the modified KS formalism with the matrix $M(s)$ is applicable to the temporally odd-number lattice. 
In fact, the periodic boundary condition for the matrix $M(s)$ is given by 
\begin{align}
M(s+N_\mu\hat{\mu})=M(s) \ \ \ \ \ \ \ (\mu=1,2,3,4), \label{OddPBC}
\end{align}
and the requirement is satisfied for all the $\mu$ on the temporally odd-number lattice 
because the spatial lattice sizes are even number and 
the matrix $M(s)$ is independent of the time component of the site $s_4$. 
Moreover, the periodic boundary condition for the staggered phase $\eta_\mu(s)$ 
is satisfied on the temporally odd-number lattice 
because the staggered phase $\eta_\mu(s)$ is also independent of the time component of the site $s_4$. 

From Eq. (\ref{MDiracM}), 
it is found that two positive modes and two negative modes appear for each eigenvalue $\lambda_n$, 
relating the spinor structure of the Dirac operator on the temporally odd-number lattice. 
Note also that the chiral symmetry guarantees the chiral partner $\gamma_5|n\rangle$ to be 
eigenmode with the eigenvalue $-i\lambda_n$. 
Thus, like the case of the even lattices, one can obtain all the eigenvalues of the Dirac operator by solving the KS Dirac eigenvalue equation (\ref{KSEigenExp}). 

In the case of the temporally odd-number lattice, 
according to the spinor structure of the Dirac operator given by Eq. (\ref{MDiracM}), 
we label the Dirac eigenstates with quantum number $I=1,2,3,4$ , namely $|n,I\rangle$. 
For each KS Dirac mode $|n)$, 
we construct these four Dirac eigenfunctions $\psi_n^{I}(s)_\alpha\equiv\langle s,\alpha|n,I\rangle$ 
using the KS Dirac eigenfunction $\chi_n(s)=\langle s|n)$, 
\begin{align}
\psi_n^{I}(s)_\alpha
=M(s)_{\alpha\beta}C_\beta^{I}\chi_n(s), \label{PsiChiOdd1}
\end{align}
where $C$ is given by Eq. (\ref{C}). 
The Dirac eigenstates $|n,I\rangle$ have the eigenvalue $i\lambda_n$ in the case of $I=1,2$ 
and have the eigenvalue $-i\lambda_n$ in the case of $I=3,4$. 
(Recall that the Dirac eigenstates with $i\lambda_n$ and the Dirac eigenstates with $-i\lambda_n$ 
appear in pairs because of chiral symmetry.) 
Substituting Eq. (\ref{C}) for Eq. (\ref{PsiChiOdd1}) one can obtain the relation 
\begin{align}
\psi_n^{I}(s)_\alpha=M(s)_{\alpha I}\chi_n(s). \label{PsiChiOdd2}
\end{align}

Next, consider rewriting the relation (\ref{RelOrig}) in terms of the KS Dirac modes. 
Taking the structure of the Dirac eigenfunction (\ref{PsiChiOdd1}) into consideration, 
Eq. (\ref{RelOrig}) should be written correctly as 
\begin{align}
 L_P =\frac{(2ai)^{N_4-1}}{12V}
\sum_{n,I}\lambda_n^{N_4-1}\langle n,I|\hat{U}_4| n,I \rangle. \label{RelOrigI}
\end{align}
Using the relation (see Appendix \ref{CalMatEleOdd})
\begin{align}
\langle n,I|\hat{U}_4| n,I \rangle=(n|\hat{U}_4| n), \label{U4DiracKSOddDiag}
\end{align}
RHS of Eq. (\ref{RelOrigI}) can be rewritten in terms of the KS Dirac modes: 
\begin{align}
&\sum_{n,I}\lambda_n^{N_4-1}\langle n,I|\hat{U}_4| n,I \rangle \nonumber\\
&\ \ \ \ =\sum_{n,I=1,2}\lambda_n^{N_4-1}\langle n,I|\hat{U}_4| n,I \rangle\nonumber\\
&\ \ \ \  \ \ +\sum_{n,I=3,4}(-\lambda_n)^{N_4-1}\langle n,I|\hat{U}_4| n,I \rangle\nonumber\\
&\ \ \ \ =\sum_{n,I=1,2,3,4}\lambda_n^{N_4-1}\langle n,I|\hat{U}_4| n,I \rangle \nonumber\\
&\ \ \ \ =\sum_{n,I=1,2,3,4}\lambda_n^{N_4-1}(n|\hat{U}_4| n) \nonumber\\
&\ \ \ \ =4\sum_{n}\lambda_n^{N_4-1}(n|\hat{U}_4| n), 
\end{align}
where $N_4-1$ is even on the temporally odd-number lattice. 
Thus, one can obtain the relation 
\begin{align}
 L_P =\frac{(2ai)^{N_4-1}}{3V}
\sum_{n}\lambda_n^{N_4-1}(n|\hat{U}_4|n) \label{RelKS}
\end{align}
using the modified KS formalism. 
Note that the (modified) KS formalism is an exact mathematical 
method for diagonalizing the Dirac operator 
and is not an approximation, 
so that Eqs.(\ref{RelOrigI}) and (\ref{RelKS}) are completely equivalent. 
Therefore, each Dirac-mode contribution to the Polyakov loop can be obtained 
by solving the eigenvalue equation of the KS Dirac operator whose dimension is $(N_c\times V)^2$ 
instead of the original Dirac operator whose dimension is $(4\times N_c\times V)^2$ 
in the case of the temporally odd-number lattice. 

Note again that we never use a specific fermion like the KS fermion here. 
We only diagonalize the Dirac operator $\slashb{D}$ 
defined by Eq.(\ref{DiracOp}) using the technique of the KS formalism, 
and obtain all the eigenvalues and the eigenfunctions of $\slashb{D}$. 
Actually, even without use of the KS formalism, 
the direct diagonalization of $\slashb{D}$ gives the same results, 
although the numerical cost is larger.

\section{Lattice QCD Numerical Analysis and Discussions}
In this section, we numerically perform SU(3) lattice QCD calculations and 
discuss the relation between confinement and chiral symmetry breaking 
based on the relation (\ref{RelKS}) connecting the Polyakov loop 
and Dirac modes on the temporally-odd number lattice. 

The SU(3) lattice QCD Monte Carlo simulations are performed 
with the standard plaquette action at the quenched level 
in both cases of confinement and deconfinement phases. 
For the confinement phase, we use a $10^3\times5$ lattice 
with $\beta\equiv\frac{2N_{\rm c}}{g^2} = 5.6$ (i.e., $a\simeq0.25$ fm), 
corresponding to $T\equiv1/(N_4 a)\simeq160$ MeV. 
For the deconfinement phase, we use $10^3\times3$ lattice 
with $\beta\equiv\frac{2N_{\rm c}}{g^2} = 5.7$ (i.e., $a\simeq0.20$ fm), 
corresponding to $T\equiv1/(N_4 a)\simeq330$ MeV. 
For each phase, we use 20 gauge configurations, 
which are taken every 500 sweeps after the thermalization of 5,000 sweeps. 

\subsection{Numerical analysis of the relation between Polyakov loop and Dirac modes}
To confirm the relation (\ref{RelKS}) numerically, 
we calculate independently LHS and RHS of the relation (\ref{RelKS}) 
and compare these values. 
A part of the numerical results in confinement and deconfinement phases 
are shown in Table \ref{RelKSConf} and Table \ref{RelKSDeconf}, respectively. 
\begin{table*}[htb]
\caption{
Numerical results for LHS and RHS of the relation (\ref{RelKS}) 
in lattice QCD with $10^3\times5$ and $\beta=5.6$ for each gauge configuration, 
where the system is in the confinement phase. 
}
\begin{tabular}{ccccccccccc} \hline \hline
Configuration No. &1&2&3&4&5&6&7&8&9&10 \\ \hline
Re$ L $
&0.00961    &-0.00161     &0.0139    &-0.00324     &0.000689   
&0.00423    &-0.00807     &-0.00918 &0.00624      &-0.00437 \\
Im$ L $
&-0.00322  &-0.00125     &-0.00438 &-0.00519     &-0.0101       
&-0.0168    &-0.00265     &-0.00683 &-0.00448     &0.00700 \\
$(3V)^{-1}\sum_{n}(2ai\lambda_n)^{N_4-1}{\rm Re}(n|\hat{U}_4|n)$ 
&0.00961    &-0.00161     &0.0139    &-0.00324     &0.000689   
&-0.00423   &-0.00807    &-0.00918 &0.00624      &-0.00437 \\
$(3V)^{-1}\sum_{n}(2ai\lambda_n)^{N_4-1}{\rm Im}(n|\hat{U}_4|n)$ 
&-0.00322  &-0.00125     &-0.00438 &-0.00519     &-0.0101     
&-0.0168    &-0.00265     &-0.00683 &-0.00448     &0.00700\\ \hline \hline
  \end{tabular}
\label{RelKSConf}
\end{table*}
\begin{table*}[htb]
\caption{
Numerical results for LHS and RHS of the relation (\ref{RelKS}) 
in lattice QCD with $10^3\times3$ and $\beta=5.7$ for each gauge configuration, 
where the system is in the deconfinement phase. 
}
\begin{tabular}{ccccccccccc} \hline \hline
Configuration No. &1&2&3&4&5&6&7&8&9&10 \\ \hline
Re$ L $
&0.316       &0.337       &0.331     &0.305      &0.313     
&0.316       &0.337       &0.300     &0.344      &0.347 \\
Im$ L $
&-0.00104  &-0.00597  &0.00723  &-0.00334 &0.00167  
&0.000120  &0.000482  &-0.00690&-0.00102 &-0.00255 \\
$(3V)^{-1}\sum_{n}(2ai\lambda_n)^{N_4-1}{\rm Re}(n|\hat{U}_4|n)$
&0.316       &0.337       &0.331     &0.305      &0.314     
&0.316       &0.337       &0.300     &0.344      &0.347 \\
$(3V)^{-1}\sum_{n}(2ai\lambda_n)^{N_4-1}{\rm Im}(n|\hat{U}_4|n)$
&-0.00104  &-0.00597   &0.00723  &-0.00334 &0.00167 
&0.000120  &0.000482   &-0.00690&-0.00102 &-0.00255 \\ \hline \hline
  \end{tabular}
\label{RelKSDeconf}
\end{table*}

From Table \ref{RelKSConf} and Table \ref{RelKSDeconf}, 
it is found that the mathematical relation (\ref{RelKS}) is exactly 
satisfied for each gauge configuration 
in both confinement and deconfinement phases, 
and this result is consistent with the analytical discussions in Sec. II. 
Then, one can discuss the relation between confinement 
and chiral symmetry breaking 
based on the relation (\ref{RelKS}) even with one gauge configuration. 
Of course, the relation is satisfied for the gauge-configuration average. 

In the deconfinement phase, the $Z_3$ center symmetry is spontaneously broken, 
and the Polyakov loop is proportional to 
$\mathrm{e}^{i\frac{2\pi}{3}j} \ (j=0,\pm 1)$ 
for each gauge configuration at the quenched level \cite{Rothe}. 
In this paper, we name the vacuum where the Polyakov loop is almost real 
($j$=0) ``real Polyakov-loop vacuum" 
and the other vacua ``$Z_3$-rotated vacua." 
At the quenched level, we have numerically confirmed that 
the relation (\ref{RelKS}) is exactly satisfied 
in the $Z_3$-rotated vacua as well as the real Polyakov-loop vacuum. 

When dynamical quarks are included, 
the real Polyakov-loop vacuum is selected as the stable vacuum, 
and the $Z_3$-rotated vacua become metastable states.
Then, the real Polyakov-loop vacuum would be more significant 
than other vacua in the deconfinement phase.
Even in full QCD, 
the mathematical relation (\ref{RelKS}) is expected to be valid, 
and we will confirm the relation and perform the 
numerical analysis in full QCD in the next study.

\subsection{Contribution from low-lying Dirac modes to Polyakov loop}
Next, we numerically confirm that low-lying Dirac modes have 
little contribution to the Polyakov loop based on the relation (\ref{RelKS}). 
This is expected from the analytical relation (\ref{RelKS}) as discussed below 
Eq.(\ref{RelOrig}), however, such a numerical analysis is also meaningful 
because the behavior of the matrix element $(n|\hat{U}_4|n)$ is nontrivial. 

Since RHS of Eq. (\ref{RelKS}) is expressed 
as a sum of the Dirac-mode contribution, 
we can calculate the Polyakov loop without low-lying Dirac-mode contribution as 
\begin{align}
(L_P)_{\rm IR\hbox{-}cut}=\frac{(2ai)^{N_4-1}}{3V}
\sum_{|\lambda_n|>\Lambda_{\rm IR}}\lambda_n^{N_4-1}(n|\hat{U}_4|n), \label{IRCutPolyakov}
\end{align}
with the infrared (IR) cutoff $\Lambda_{\rm IR}$ for Dirac eigenvalue. 
The chiral condensate $\langle \bar{q}q \rangle$ is expressed as 
\begin{eqnarray}
 \langle \bar qq\rangle
&=&-\frac{1}{V}{\rm Tr}_{c,\gamma}\frac{1}{\slashb D+m}
=-\frac{1}{V}\sum_n\frac{1}{i\lambda_n+m} \nonumber\\
&=&-\frac{1}{V}\left(\sum_{\lambda_n>0} \frac{2m}{\lambda_n^2+m^2}
+\frac{\nu}{m}\right), 
\label{qbarq}
\end{eqnarray}
where $m$ is the current quark mass 
and $\nu$ the total number of zero modes of $\slashb D$. 

\begin{figure}[h]
\begin{center}
\includegraphics[scale=0.5]{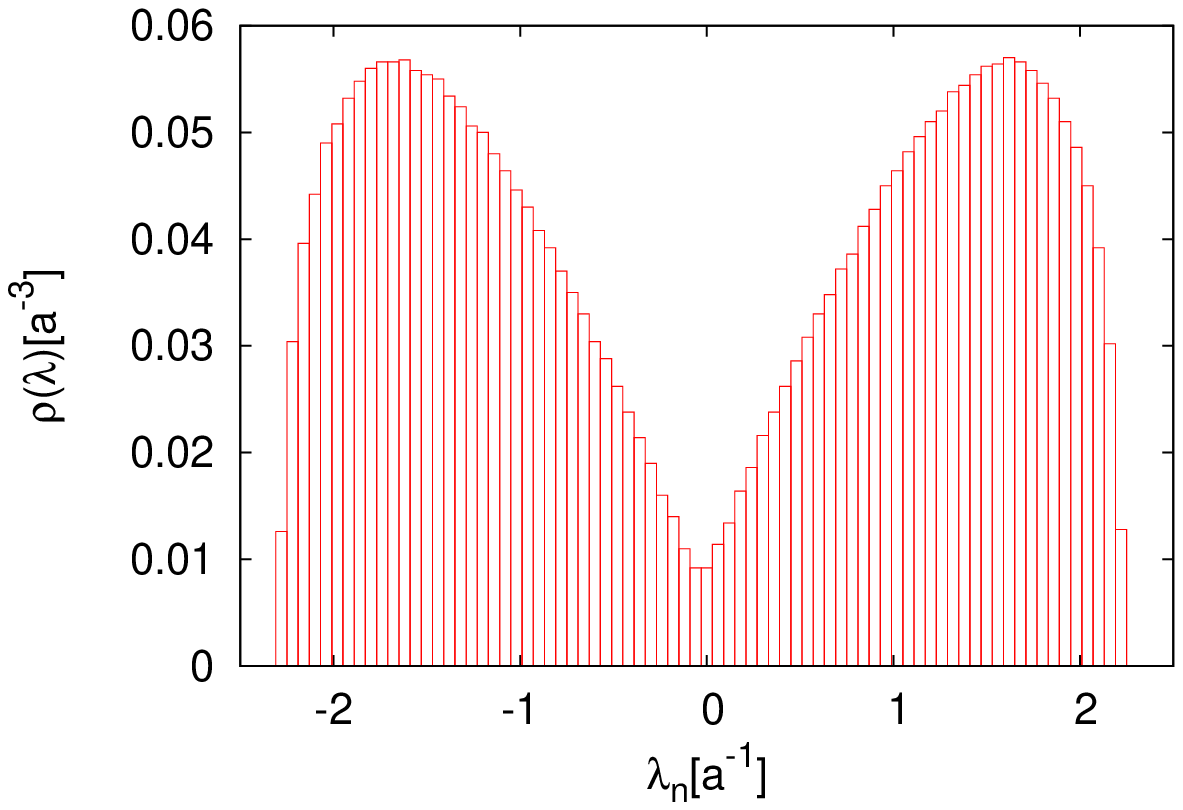}
\includegraphics[scale=0.5]{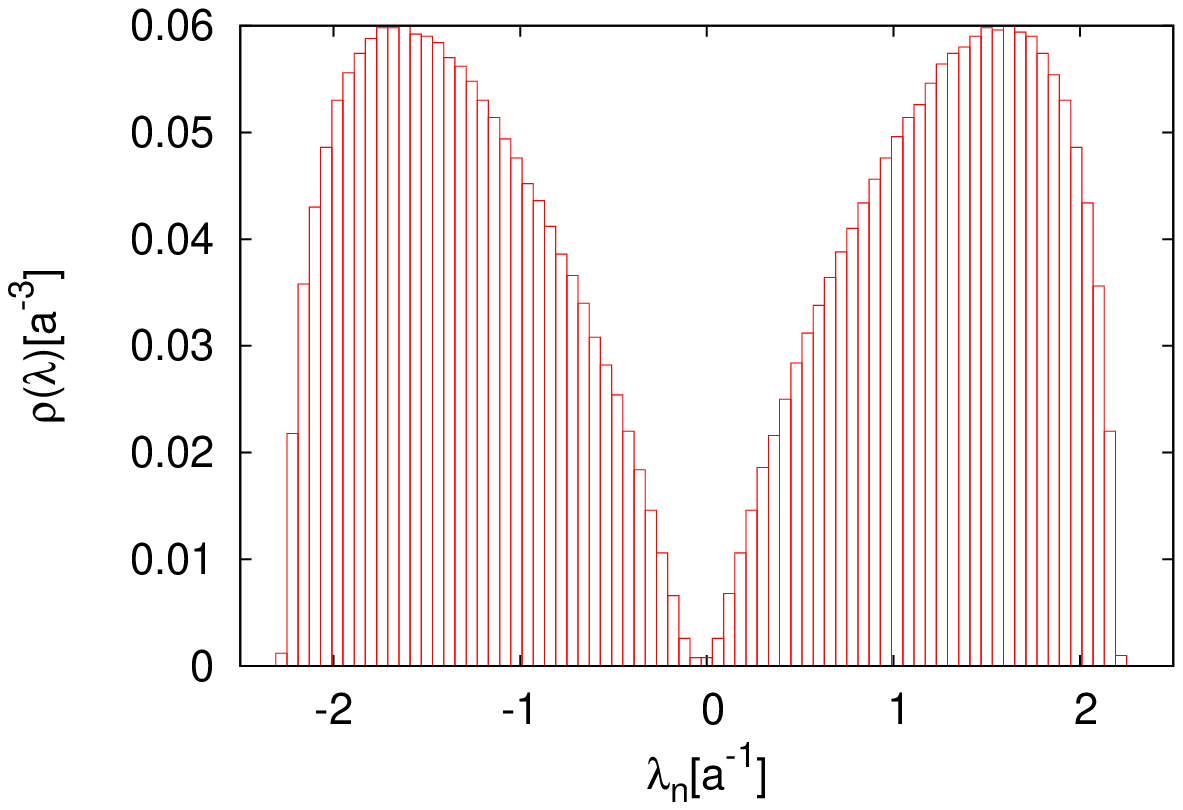}
\caption{
The lattice QCD result of the Dirac eigenvalue distribution $\rho(\lambda)$ 
in confinement and deconfinement phases in the lattice unit. 
The upper figure shows $\rho(\lambda)$ in the confinement phase 
on $10^3\times5$ lattice with $\beta\equiv\frac{2N_{\rm c}}{g^2} = 5.6$ (i.e., $a\simeq0.25$ fm). 
The lower figure shows $\rho(\lambda)$ in the deconfinement phase 
on $10^3\times5$ lattice with $\beta\equiv\frac{2N_{\rm c}}{g^2} = 6.0$ (i.e., $a\simeq0.10$ fm). 
}
\label{DiracEigenDist}
\end{center}
\end{figure}

We show the lattice QCD result of 
the Dirac eigenvalue distribution $\rho(\lambda)$ 
in confinement and deconfinement phases in Fig. \ref{DiracEigenDist}. 
In the deconfinement phase, 
the number of low-lying Dirac modes is significantly reduced 
and $\rho(\lambda=0) \simeq 0$, which means that 
the chiral condensate is almost zero and the chiral symmetry is restored. 
Then, in the deconfinement phase, 
it may be less interesting to investigate 
the effect of low-lying Dirac modes to the Polyakov loop, 
because low-lying Dirac modes are almost absent. 

The chiral condensate after the removal of 
contribution from the low-lying Dirac modes below 
IR cutoff $\Lambda_{\rm IR}$ is expressed as 
\begin{align}
\langle \bar{q}q\rangle_{\Lambda_{\rm IR}}
=-\frac{1}{V}\sum_{\lambda_n\geq\Lambda_{\rm IR}}\frac{2m}{\lambda_n^2+m^2}
\end{align}
In this paper, we take the IR cutoff of $\Lambda_{\rm IR}\simeq0.4 {\rm GeV}$.
In the confined phase, this IR Dirac-mode cut leads to 
\begin{eqnarray}
\frac{\langle \bar{q}q\rangle_{\Lambda_{\rm IR}}}
{\langle \bar{q}q\rangle} \simeq 0.02 \label{qqbarIRcutRatio}
\end{eqnarray}
and almost chiral-symmetry restoration 
in the case of physical current-quark mass, $m\simeq5 {\rm MeV}$ \cite{GIS}. 

A part of the numerical results 
for $L_P$ and $(L_P)_{\rm IR\hbox{-}cut}$ 
with the IR cutoff of $\Lambda_{\rm IR}\simeq0.4 {\rm GeV}$ 
in both confinement and deconfinement phases 
are shown in Table \ref{LowConf} and Table \ref{LowDeconf}, respectively. 

\begin{table*}[htb]
\caption{
Numerical results for $L_P$ and 
$(L_P)_{\rm IR\hbox{-}cut}$ 
in lattice QCD with $10^3\times5$ and $\beta=5.6$ 
for each gauge configuration, 
where the system is in the confinement phase.
}
  \begin{tabular}{ccccccccccc} \hline \hline
Configuration No. &1&2&3&4&5&6&7&8&9&10 \\ \hline
Re$L_P$             
&0.00961 &-0.00161&0.0139    &-0.00324&0.000689
&0.00423 &-0.00807&-0.00918&0.00624  &-0.00437\\ 
Im$L_P$             
&-0.00322&-0.00125&-0.00438&-0.00519&-0.0101  
&-0.0168 &-0.00265&-0.00683 &-0.00448&0.00700 \\
Re$(L_P)_{\rm IR\hbox{-}cut}$ 
&0.00961 &-0.00160&0.0139    &-0.00325&0.000706
&0.00422&-0.00807 &-0.00918 &0.00624 &-0.00436\\ 
Im$(L_P)_{\rm IR\hbox{-}cut}$ 
&-0.00321&-0.00125&-0.00437&-0.00520&-0.0101  
&-0.0168&-0.00264  &-0.00682&-0.00448&0.00698 \\ \hline \hline 
  \end{tabular}
\label{LowConf}
\end{table*}
\begin{table*}[htb]
\caption{
Numerical results for $L_P$ and 
$(L_P)_{\rm IR\hbox{-}cut}$ 
in lattice QCD with $10^3\times3$ and $\beta=5.7$ 
for each gauge configuration, 
where the system is in the deconfinement phase.
}
  \begin{tabular}{ccccccccccc} \hline \hline
Configuration No. &1&2&3&4&5&6&7&8&9&10 \\ \hline
Re$L_P$             
&0.316     &0.337       &0.331     &0.305     &0.314   
&0.316     &0.337       &0.300     &0.344     &0.347 \\
Im$L_P$              
&-0.00104&-0.00597  &0.00723  &-0.00334&0.00167
&0.000120&0.0000482&-0.00690&-0.00102 &-0.00255 \\
Re$(L_P)_{\rm IR\hbox{-}cut}$ 
&0.319     &0.340       &0.334     &0.307     &0.317   
&0.319      &0.340      &0.303     &0.347     &0.350 \\
Im$(L_P)_{\rm IR\hbox{-}cut}$ 
&-0.00103&-0.00597  &0.00724  &-0.00333&0.00167
&0.000121 &0.0000475&-0.000691&-0.00102&-0.00256\\ \hline \hline
  \end{tabular}
\label{LowDeconf}
\end{table*}
From Table \ref{LowConf} and Table \ref{LowDeconf}, 
it is found that 
$L_P\simeq(L_P)_{\rm IR\hbox{-}cut}$  
is almost satisfied for each gauge configuration 
in both confinement and deconfinement phases. 
In the deconfinement phase, 
we have confirmed that 
$L_P\simeq(L_P)_{\rm IR\hbox{-}cut}$ is satisfied 
for both real Polyakov-loop vacuum and 
$Z_3$-rotated vacua.
Thus, the configuration average $\langle L_P\rangle\simeq\langle (L_P)_{\rm IR\hbox{-}cut}\rangle$ 
is of course almost satisfied. 
Therefore, the low-lying Dirac modes have little contribution to the Polyakov loop 
and are not essential for confinement. 
From Eq. (\ref{qqbarIRcutRatio}), however, the low-lying Dirac modes 
below the IR cutoff $|\lambda_n|<\Lambda_{\rm IR}\simeq0.4 {\rm GeV}$ are essential 
for chiral symmetry breaking. 
Thus, we conclude that there is no one-to-one 
correspondence between confinement and chiral symmetry breaking.
This result is consistent with the previous numerical lattice analysis that 
the confinement properties such as the Polyakov loop and the string tension, or confinement force, are almost unchanged by removing low-lying Dirac modes from QCD vacuum \cite{GIS}. 

\subsection{New ``positive/negative symmetry" on Dirac matrix element 
in confinement phase}
Since Eq. (\ref{RelKS}) is the Dirac spectral expression of the Polyakov loop, 
one can investigate the contribution from each Dirac mode to the Polyakov loop. 
We calculate the matrix element $(n|\hat{U}_4|n)$ 
and each Dirac-mode contribution $\lambda_n^{N_4-1}(n|\hat{U}_4|n)$ 
in both confinement and deconfinement phases. 
The Polyakov loop is obtained by multiplying 
the sum of each Dirac-mode contribution $\sum_n \lambda_n^{N_4-1}(n|\hat{U}_4|n)$ 
by the overall factor $(2ai)^{N_4-1}/(3V)$ in Eq. (\ref{RelKS}). 

\subsubsection{Confinement phase case}
\begin{figure}[h]
\begin{center}
\includegraphics[scale=0.5]{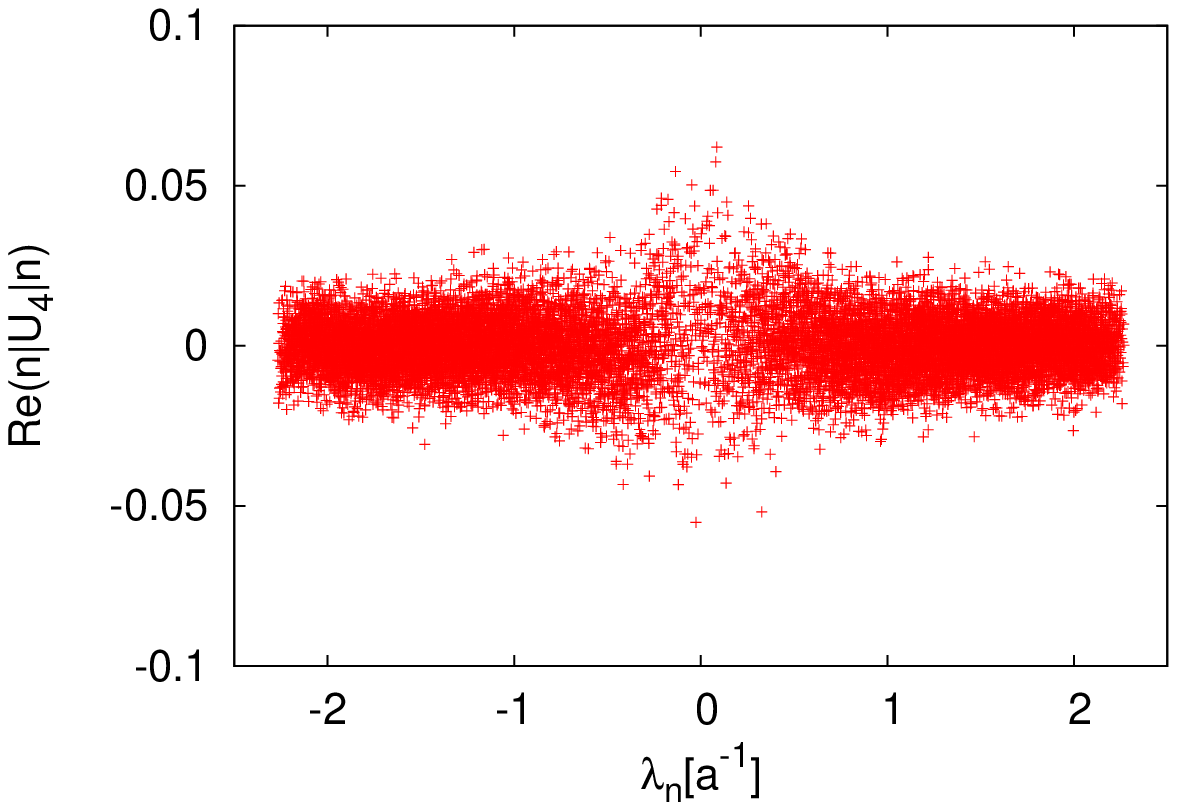}
\includegraphics[scale=0.5]{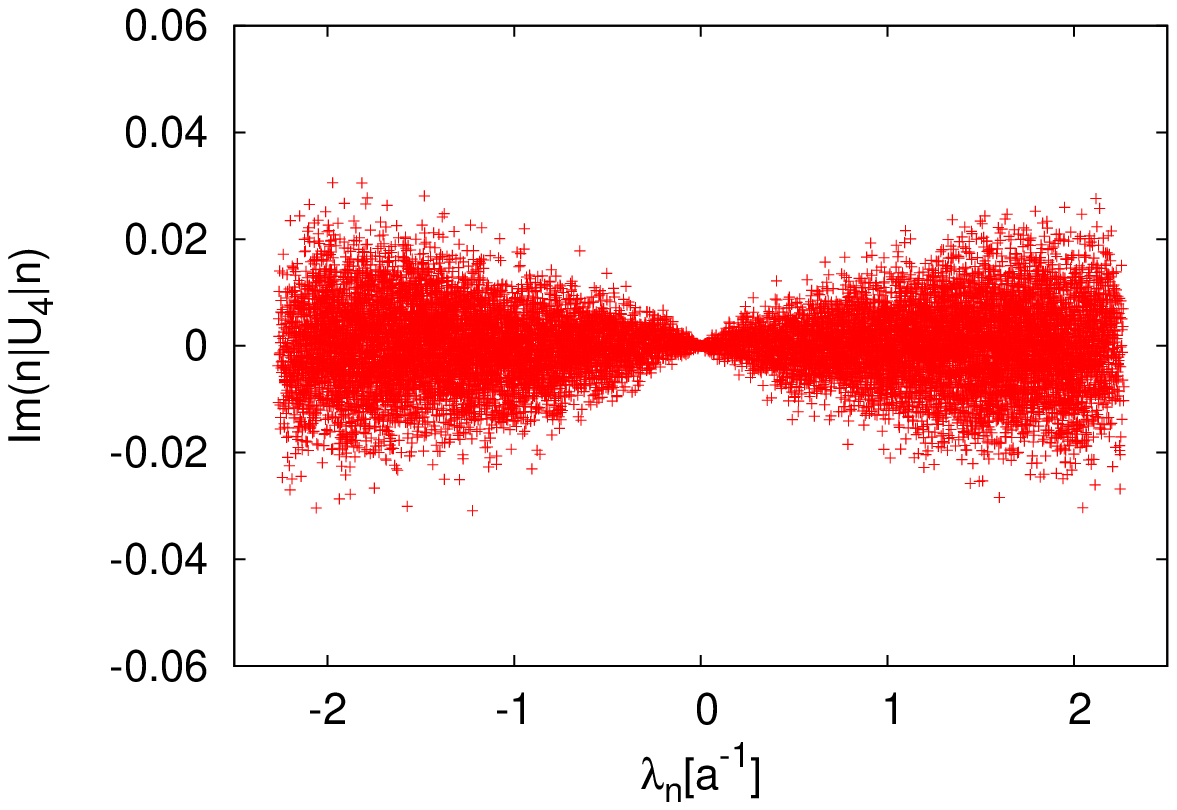}
\caption{
The real part Re$(n|\hat{U}_4|n)$ 
and the imaginary part Im$(n|\hat{U}_4|n)$ of the matrix element 
in the confinement phase, 
plotted against the Dirac eigenvalue $\lambda_n$ 
in the lattice unit at $\beta=5.6$ on $10^3\times 5$. 
There is the positive/negative symmetry. 
}
\label{MatEleConf}
\end{center}
\end{figure}

\begin{figure}[h]
\begin{center}
\includegraphics[scale=0.5]{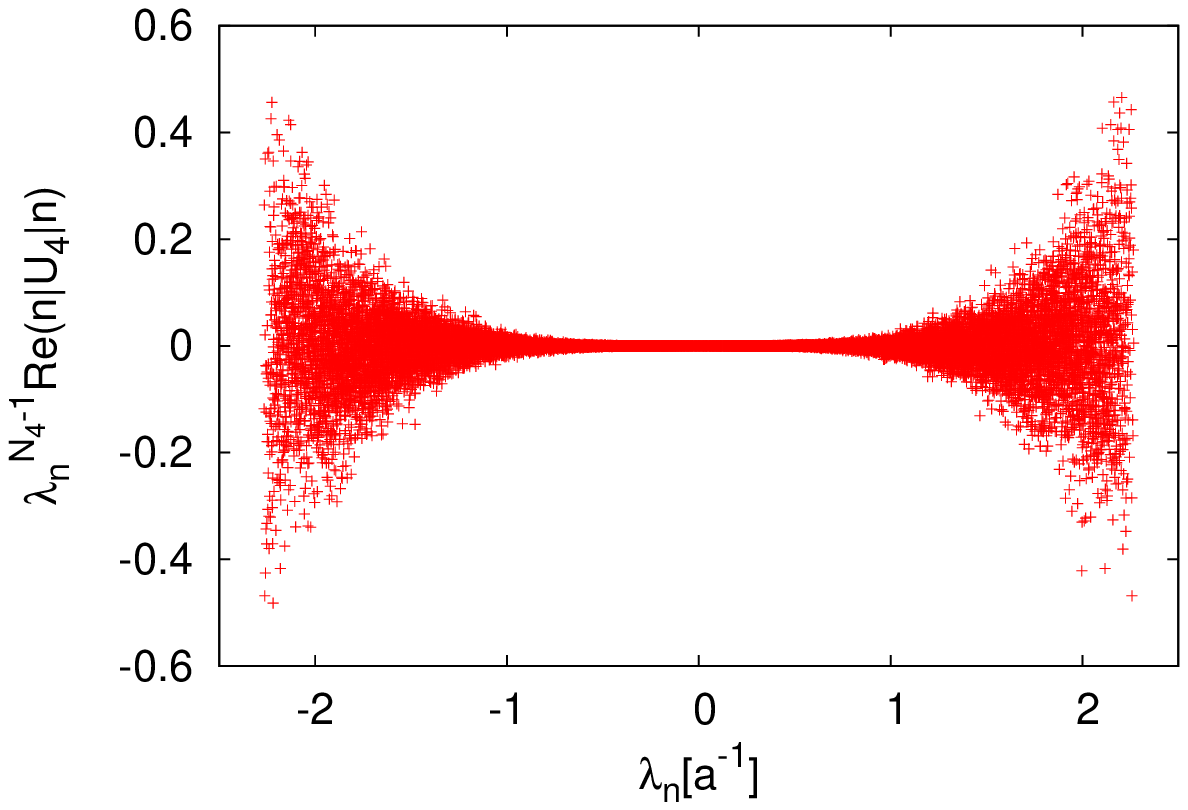}
\includegraphics[scale=0.5]{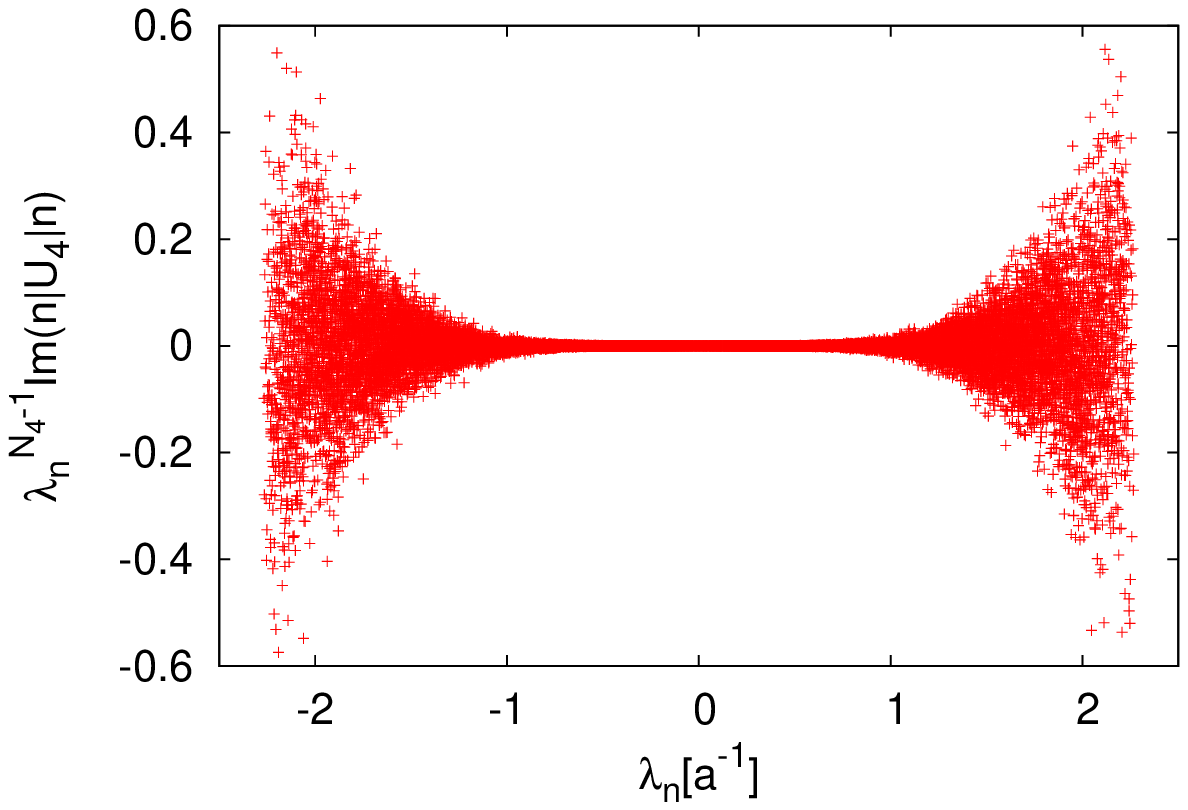}
\caption{
Each Dirac-mode contribution to the Polyakov loop, 
$\lambda_n^{N_4-1}{\rm Re}(n|\hat{U}_4|n)$ 
and $\lambda_n^{N_4-1}{\rm Im}(n|\hat{U}_4|n)$ 
in the confinement phase, 
plotted against the Dirac eigenvalue $\lambda_n$ 
in the lattice unit 
at $\beta=5.6$ on $10^3\times5$. 
There is the positive/negative symmetry. 
}
\label{ContConf}
\end{center}
\end{figure}

Figure \ref{MatEleConf} shows the numerical results for the matrix elements 
Re$(n|\hat{U}_4|n)$ and Im$(n|\hat{U}_4|n)$ 
plotted against Dirac eigenvalues $\lambda_n$ in the lattice unit 
for one gauge configuration in the confinement phase. 
Figure~\ref{ContConf} shows each Dirac-mode contribution to the Polyakov loop 
$\lambda_n^{N_4-1}{\rm Re}(n|\hat{U}_4|n)$ and $\lambda_n^{N_4-1}{\rm Im}(n|\hat{U}_4|n)$ 
plotted against Dirac eigenvalues $\lambda_n$ in the lattice unit. 
In the confinement phase, 
the real part of the matrix element ${\rm Re}(n|\hat{U}_4|n)$ 
is generally nonzero 
in the whole region and is not small in low-lying Dirac-mode region 
from Fig. \ref{MatEleConf}. 
However, the Dirac-mode contribution to the Polyakov loop, $\lambda_n^{N_4-1}{\rm Re}(n|\hat{U}_4|n)$, 
is small in low-lying Dirac-mode region because of the damping factor $\lambda_n^{N_4-1}$ 
from Fig. \ref{ContConf}. 
Thus, the damping factor $\lambda_n^{N_4-1}$ has an essential role in 
Eq. (\ref{RelKS}). 

On the other hand, from Fig. \ref{MatEleConf}, 
the imaginary part ${\rm Im}(n|\hat{U}_4|n)$ of the matrix element 
is relatively small in low-lying Dirac-mode region, 
in comparison with ${\rm Re}(n|\hat{U}_4|n)$. 
In any case, $\lambda_n^{N_4-1}{\rm Im}(n|\hat{U}_4|n)$ 
is small in low-lying Dirac-mode region, 
as shown in Fig. \ref{ContConf}. 

Remarkably, as shown in Fig. \ref{MatEleConf}, 
there is a new symmetry of ``positive/negative symmetry" in the confinement phase 
for the distribution of Dirac-mode matrix element $(n|\hat{U}_4|n)$, 
i.e., ${\rm Re}(n|\hat{U}_4|n)$ and ${\rm Im}(n|\hat{U}_4|n)$. 
Then, 
the distribution of each Dirac-mode contribution to the Polyakov loop, $\lambda_n^{N_4-1}(n|\hat{U}_4|n)$,  has the same symmetry. 
Since the Polyakov loop is proportional to the total sum of 
each Dirac-mode contribution, $\sum_n \lambda_n^{N_4-1}(n|\hat{U}_4|n)$, 
this new symmetry leads to the zero value of the Polyakov loop, i.e., $\langle L_P\rangle=0$, 
in the confinement phase. 
Moreover, the contribution to the Polyakov loop from arbitrary Dirac-mode region $\Lambda_1\leq\lambda_n\leq\Lambda_2$ is zero 
due to the symmetry in the confinement phase: 
\begin{eqnarray}
\sum_{\Lambda_1\leq\lambda_n\leq\Lambda_2}\lambda_n^{N_4-1}(n|\hat{U}_4| n)
=0 
\ \ ({\rm confinement \ phase})
. 
\end{eqnarray}
This behavior in the confinement phase is consistent with the previous works \cite{GIS}. 

Note that the distribution of 
the matrix elements $(n|\hat{U}_4|n)$ 
is not statistical fluctuation on the gauge ensemble 
because the results shown here are for one configuration. 
We find the same behavior for other gauge configurations. 

As for the $N_4$ dependence of the matrix element $(n|\hat{U}_4|n)$ in the confinement phase, 
we find almost the same results that 
there is the positive/negative symmetry 
and low-lying Dirac modes have little contribution to the Polyakov loop. 

\subsubsection{Deconfinement phase case}
Since the deconfinement phase does not have 
confinement and chiral symmetry breaking, 
it may be less interesting to consider their relation there. 
In the deconfinement phase, the $Z_3$ center symmetry is 
spontaneously broken, and there appear three types of vacua 
corresponding to the Polyakov loop 
proportional to $\mathrm{e}^{i\frac{2\pi}{3}j} \ (j=0,\pm 1)$, 
while the confinement phase has a unique vacuum 
of $L_P \simeq 0$ on the $Z_3$ symmetry.
Here, we mainly consider the real Polyakov-loop vacuum, 
since it is selected as the stable vacuum 
when dynamical quarks are included.

\begin{figure}[h]
\begin{center}
\includegraphics[scale=0.5]{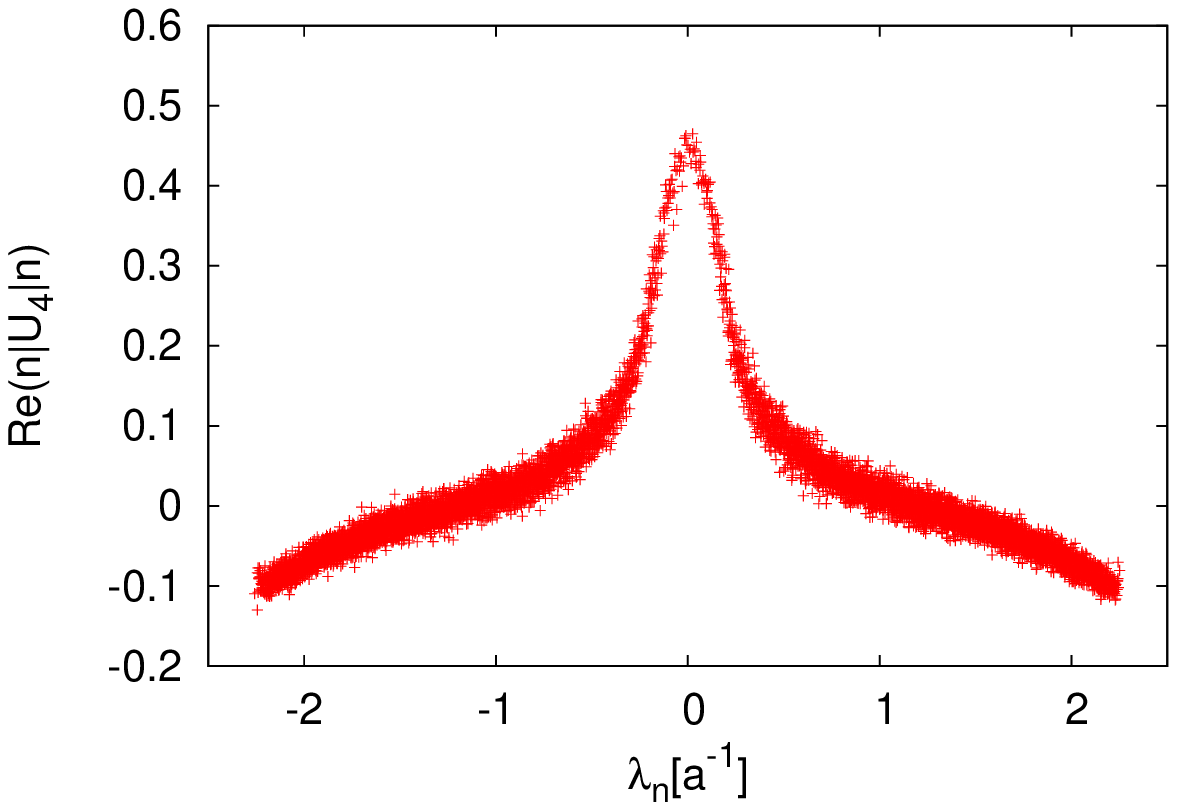}
\includegraphics[scale=0.5]{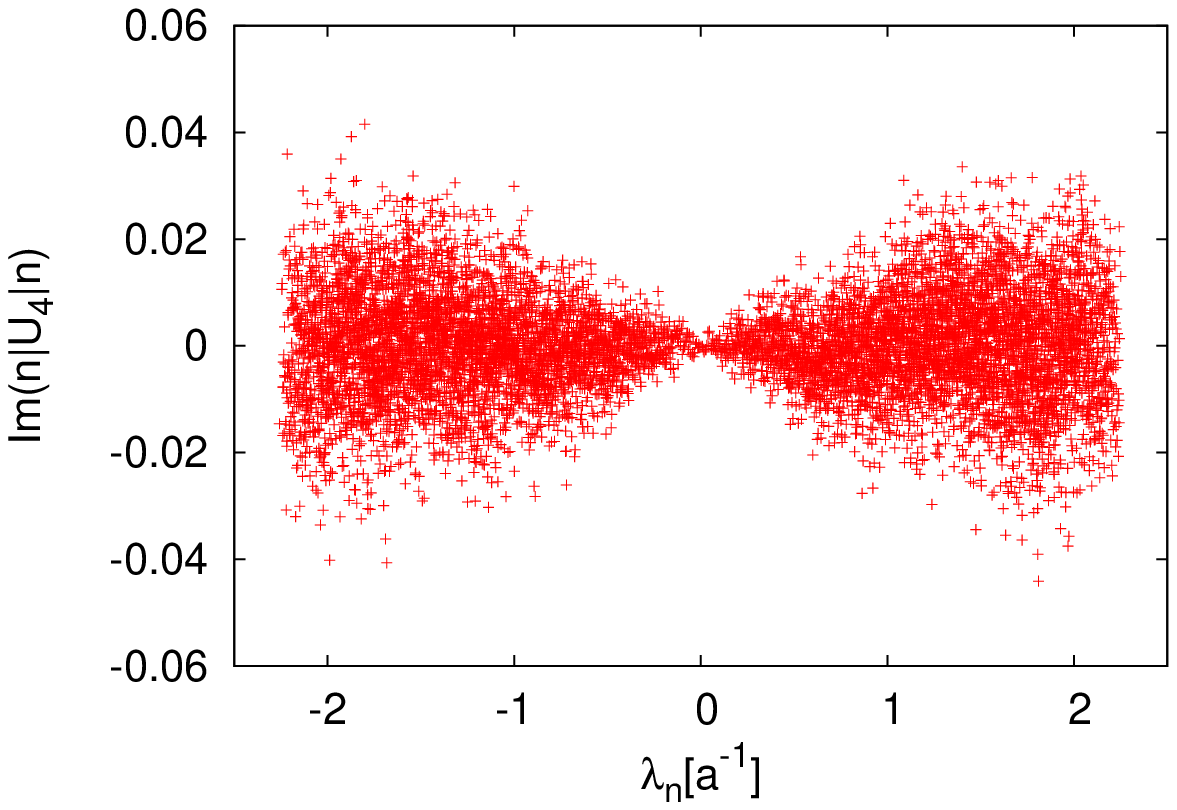}
\caption{
The real part Re$(n|\hat{U}_4|n)$ 
and the imaginary part Im$(n|\hat{U}_4|n)$ of the matrix element 
in the deconfinement phase with real Polyakov loop, 
plotted against the Dirac eigenvalue $\lambda_n$
in the lattice unit 
at $\beta=5.7$ on $10^3\times3$. 
}
\label{MatEleDeconf}
\end{center}
\end{figure}

\begin{figure}[h]
\begin{center}
\includegraphics[scale=0.5]{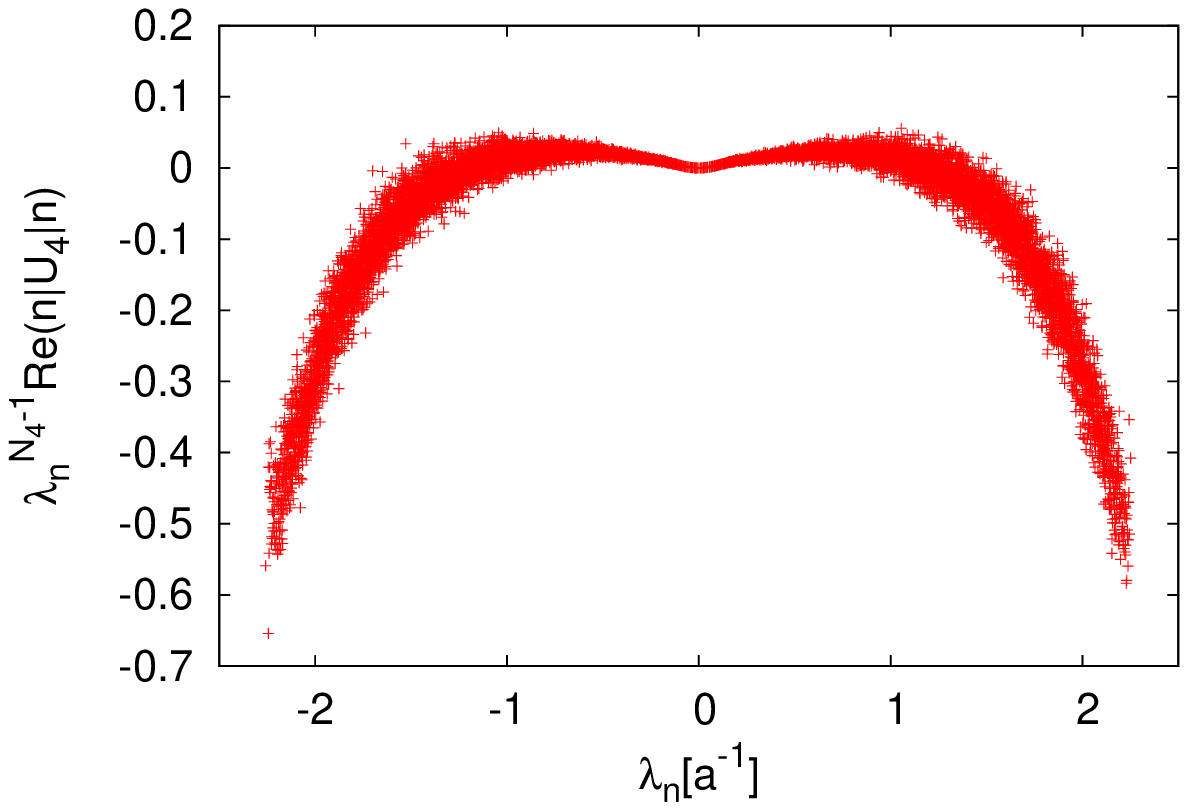}
\includegraphics[scale=0.5]{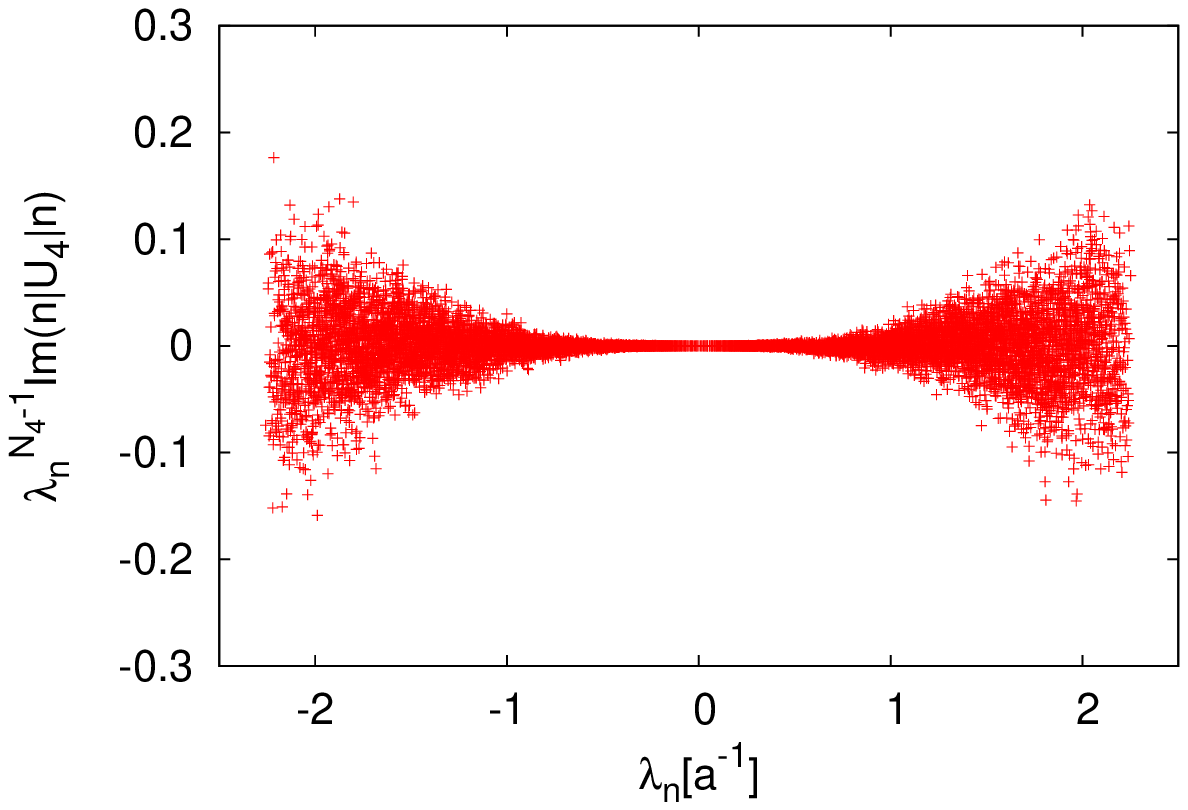}
\caption{
Each Dirac-mode contribution to the Polyakov loop, 
$\lambda_n^{N_4-1}{\rm Re}(n|\hat{U}_4|n)$ and 
$\lambda_n^{N_4-1}{\rm Im}(n|\hat{U}_4|n)$, 
in the deconfinement phase with real Polyakov loop, 
plotted against the Dirac eigenvalue $\lambda_n$ in the lattice unit 
at $\beta=5.7$ on $10^3\times3$. 
}
\label{ContDeconf}
\end{center}
\end{figure}

We show in Figs. \ref{MatEleDeconf} and \ref{ContDeconf} 
the matrix elements $(n|\hat{U}_4|n)$ 
and each Dirac-mode contribution $\lambda_n^{N_4-1}{\rm Re}(n|\hat{U}_4|n)$ 
in the deconfinement phase with real Polyakov loop, 
plotted against the Dirac eigenvalue $\lambda_n$, in quenched lattice QCD. 
The imaginary part ${\rm Im}(n|\hat{U}_4|n)$ of the matrix element 
shows the same behavior as the case of the confinement phase, 
where the gauge-configuration average of the Polyakov loop is zero. 
(Compare Fig. \ref{MatEleConf}(b) and Fig. \ref{MatEleDeconf}(b).) 
Then, we consider only the results for real part of these quantities 
in the deconfinement phase. 
Like the case of the confinement phase, 
we show the results for one gauge configuration 
since the results are almost the same for the other configuration. 

From Fig. \ref{MatEleDeconf}, the real part of the matrix element, ${\rm Re}(n|\hat{U}_4|n)$, 
has a peak in low-lying Dirac-mode region. 
However, from Fig. \ref{ContDeconf}, each Dirac-mode contribution 
$\lambda_n^{N_4-1}{\rm Re}(n|\hat{U}_4|n)$ is 
relatively small in low-lying Dirac-mode region because of the damping factor $\lambda_n^{N_4-1}$ 
like the case of confinement phase. 
The Dirac-mode contribution $\lambda_n^{N_4-1}{\rm Re}(n|\hat{U}_4|n)$ 
takes a negative value for most regions of $\lambda_n$, as shown in Fig. \ref{ContDeconf}(a). 
This is consistent with the positive value of the Polyakov loop and $N_4=3$, 
considering the overall factor $(2ai)^{N_4-1}/(3V)$ in Eq. (\ref{RelKS}). 
More quantitatively, 
only high-lying Dirac modes have contribution to the nonzero value of the Polyakov loop 
from Fig. \ref{ContDeconf}. 

In the deconfinement phase, there is no more positive/negative symmetry for 
the distributions of the matrix element $(n|\hat{U}_4|n)$ 
and each Dirac-mode contribution $\lambda_n^{N_4-1}(n|\hat{U}_4|n)$, 
unlike the case of the confinement phase with the symmetry. 
The Polyakov loop is nonzero 
because of the asymmetry in the distribution of the matrix element and each Dirac-mode contribution, 
while the Polyakov loop in the confinement phase is zero because of the symmetry. 
Thus, the appearance of the positive/negative symmetry on the matrix element $(n|\hat{U}_4|n)$ 
is strongly related to the deconfinement phase transition. 
This behavior is similar to the $Z_3$ center symmetry, 
which is not broken in the confinement phase and is broken in the deconfinement phase 
at the quenched level. 
Therefore, it is interesting to investigate the relation between the new positive/negative symmetry 
and the $Z_3$ center symmetry. 

\begin{figure}[h]
\begin{center}
\includegraphics[scale=0.5]{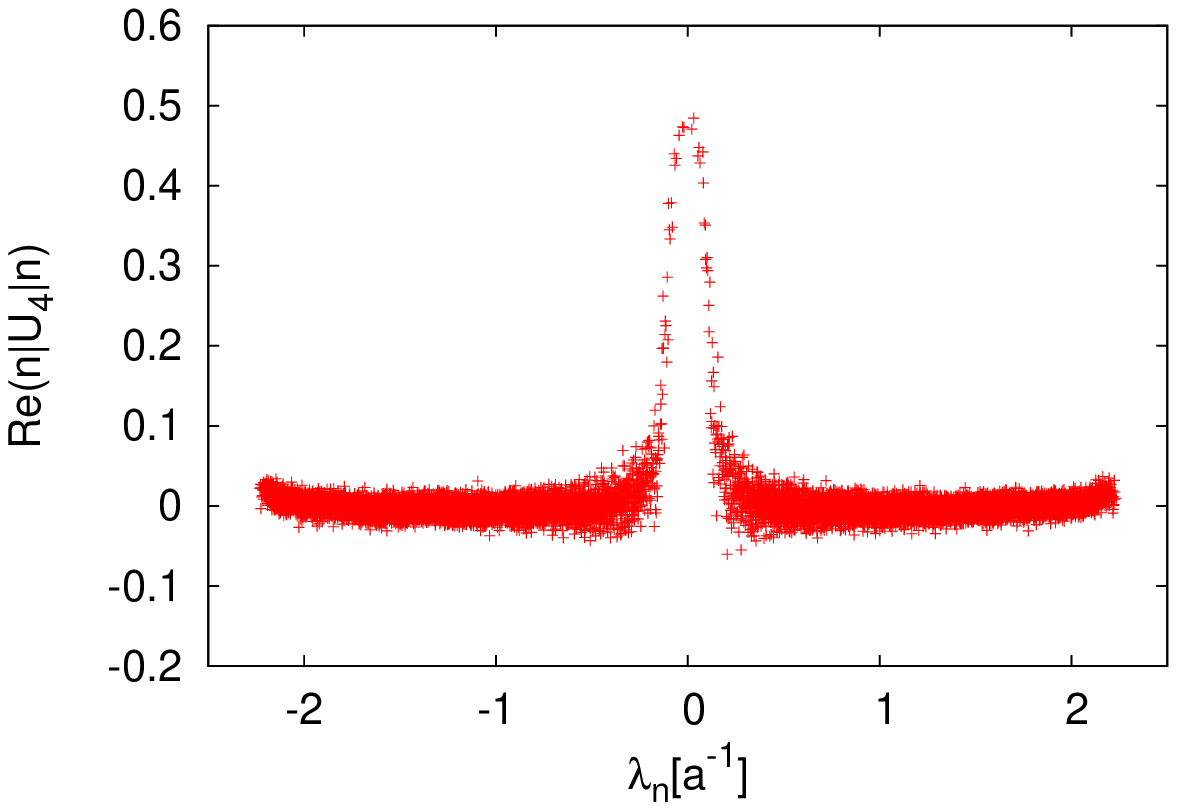}
\includegraphics[scale=0.5]{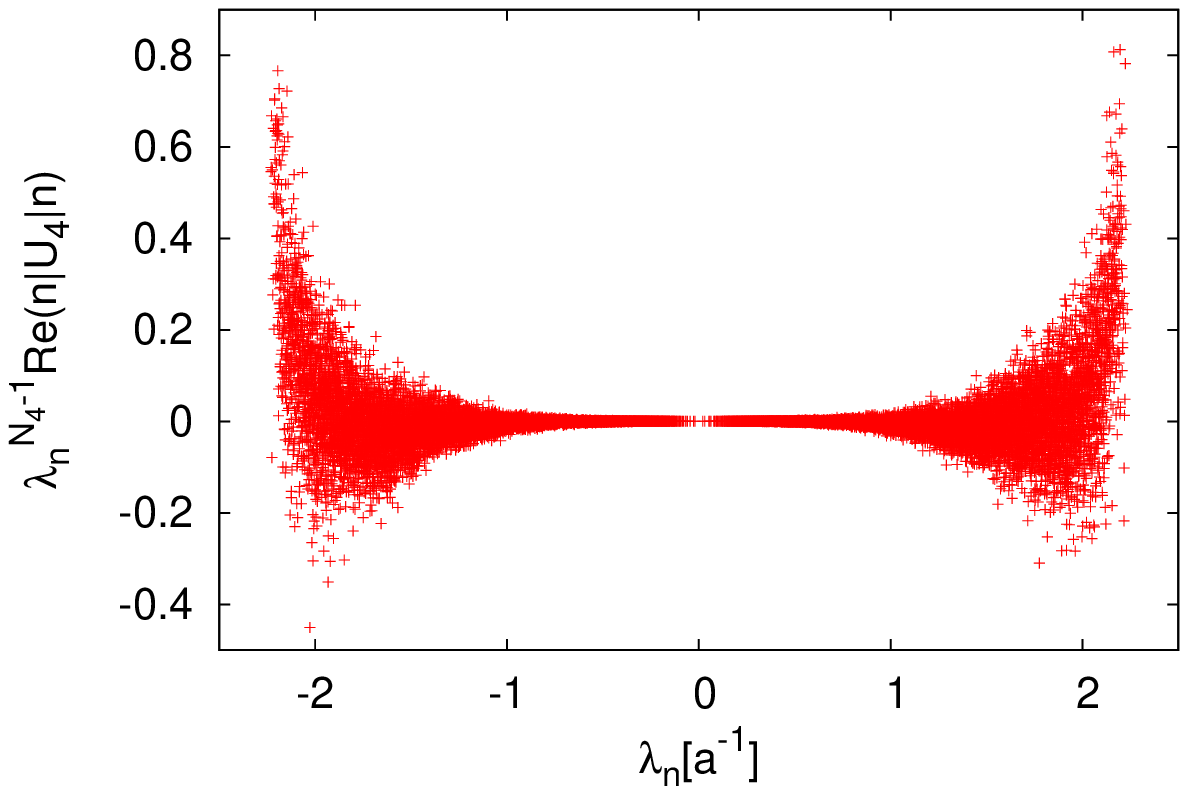}
\caption{
The real part of the matrix element Re$(n|\hat{U}_4|n)$ 
and each Dirac-mode contribution to the Polyakov loop 
$\lambda_n^{N_4-1}{\rm Re}(n|\hat{U}_4|n)$ in 
the deconfinement phase with real Polyakov loop, 
plotted against the Dirac eigenvalue $\lambda_n$ in the lattice unit 
at $\beta=6.0$ on $10^3\times5$. 
The sign of $\lambda_n^{N_4-1}{\rm Re}(n|\hat{U}_4|n)$ 
is different from Fig. \ref{ContDeconf} 
due to the overall factor $(2ai)^{N_4-1}/(3V)$ in Eq.(\ref{RelKS}). 
}
\label{MatEleandContDeconf}
\end{center}
\end{figure}

Next, we consider $N_4$ dependence of the matrix element $(n|\hat{U}_4|n)$ 
in the deconfinement phase with real Polyakov loop. 
We numerically confirm that the relation (\ref{RelKS}) is satisfied exactly 
and the contribution from the low-lying Dirac modes 
to the Polyakov loop is negligible regardless of 
the temporal lattice size $N_4$. 
Figure \ref{MatEleandContDeconf} shows 
results for the $10^3\times5$ lattice with $\beta\equiv\frac{2N_{\rm c}}{g^2}=6.0$ 
(i.e., $a\simeq0.10$ fm), corresponding to $T\equiv1/(N_4a)\simeq400$ MeV. 
Since the Polyakov loop is real in our calculation, 
we show only the real part of the matrix element 
and each Dirac-mode contribution 
in Fig. \ref{MatEleandContDeconf}. 
There are some points in common between the $N_4=3$ and the $N_4=5$ cases. 
We find again no positive/negative symmetry 
and the real part of the matrix element Re$(n|\hat{U}_4|n)$ has a peak in low-lying Dirac-mode region 
and low-lying Dirac modes have little contribution to the Polyakov loop 
because of the damping factor $\lambda_n^{N_4-1}$. 
However, there is difference in the shape of the distribution of the matrix element Re$(n|\hat{U}_4|n)$ 
between Figs. \ref{MatEleDeconf} and \ref{MatEleandContDeconf}. 
The total sum of the Dirac-mode contribution $\lambda_n^{N_4-1}{\rm Re}(n|\hat{U}_4|n)$ is positive, as shown in Fig. \ref{MatEleandContDeconf}(b). 
This is consistent with the positive value of the Polyakov loop and $N_4=5$, 
considering the overall factor $(2ai)^{N_4-1}/(3V)$ in Eq. (\ref{RelKS}). 
In any case, independent of lattice size, 
the positive/negative symmetry and the damping factor 
$\lambda_n^{N_4-1}$ are important for the behavior of 
the Polyakov loop and the low-lying Dirac-mode contribution. 

\begin{figure}[h]
\begin{center}
\includegraphics[scale=0.5]{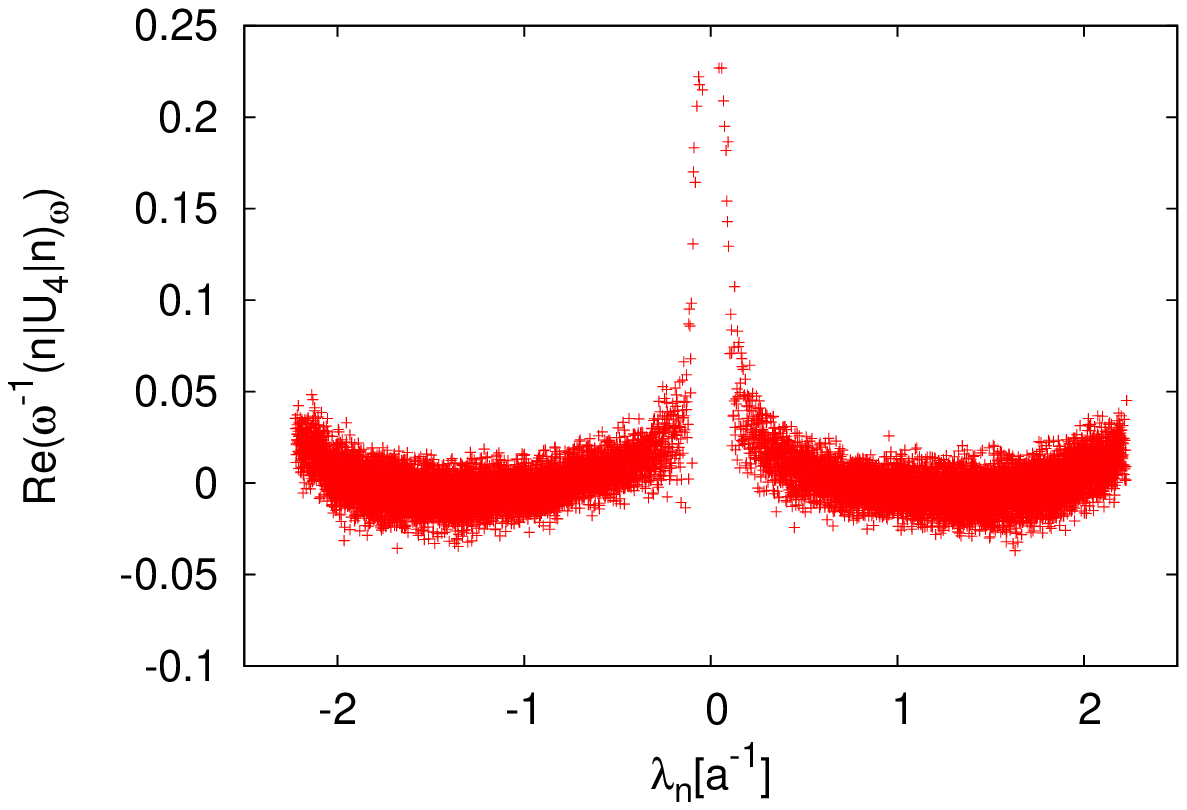}
\includegraphics[scale=0.5]{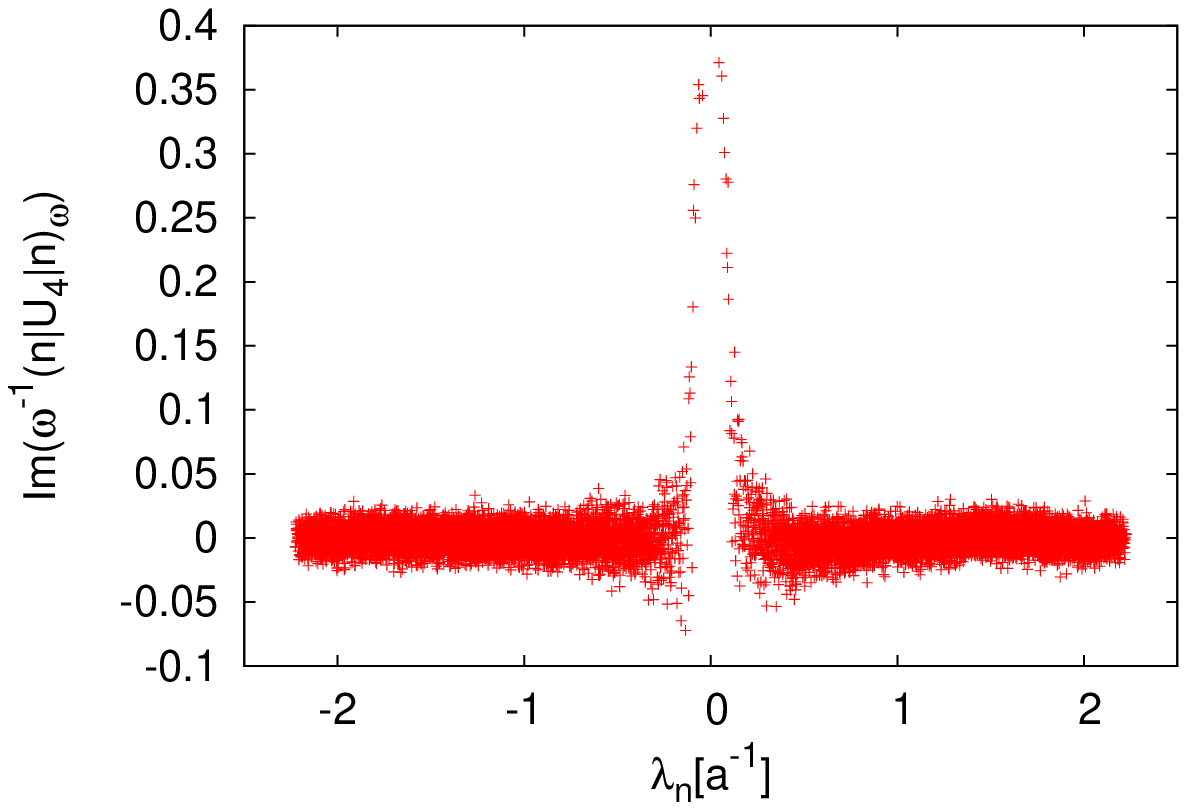}
\caption{
The longitudinal matrix element Re$(\omega^{-1}(n|\hat{U}_4|n)_\omega)$
and the transverse matrix element Im$(\omega^{-1}(n|\hat{U}_4|n)_\omega)$ 
in the $Z_3$-rotated vacuum by $\omega$ 
in the deconfinement phase, plotted against Dirac eigenvalues $\lambda_n$ 
in the lattice unit at $\beta=6.0$ on $10^3\times5$. 
}
\label{MatEleandContDeconfOmega}
\end{center}
\end{figure}

Also, we investigate the $Z_3$-rotated vacuum 
in the deconfinement phase and the Dirac modes there, 
although this vacuum is metastable and less significant 
when dynamical quarks are included. 
The $Z_3$-rotated vacuum can be practically generated 
by changing the initial condition in our Monte Carlo simulation. 
%
Using $Z_3$ factors $\omega\equiv\mathrm{e}^{2\pi i/3}$ 
and $\omega^2=\mathrm{e}^{4\pi i/3}$, 
we denote the matrix element in the $\omega$-rotated configuration 
by $(n|\hat{U}_4|n)_\omega$. 
For the comparison between the matrix element $(n|\hat{U}_4|n)_\omega$ 
in the $\omega$-rotated configuration 
and $(n|\hat{U}_4|n)$ in the real Polyakov-loop configuration, 
we define 
longitudinal and transverse matrix elements, 
${\rm Re}(\omega^{-1}(n|\hat{U}_4|n)_\omega)$ 
and ${\rm Im}(\omega^{-1}(n|\hat{U}_4|n)_\omega)$, 
for the $\omega$-rotated configuration.
The longitudinal and transverse matrix elements correspond to 
${\rm Re}(n|\hat{U}_4|n)$ and ${\rm Im}(n|\hat{U}_4|n)$ 
in the real Polyakov-loop configuration, respectively. 
%
Figure \ref{MatEleandContDeconfOmega} shows 
the matrix elements in the $Z_3$-rotated vacuum by $\omega$ 
on the $10^3\times5$ lattice at $\beta\equiv\frac{2N_{\rm c}}{g^2}=6.0$ 
(i.e., $a\simeq0.10$ fm), corresponding to $T\equiv1/(N_4a)\simeq400$ MeV. 
There is no positive/negative symmetry in the distribution of 
the longitudinal matrix elements 
${\rm Re}(\omega^{-1}(n|\hat{U}_4|n)_\omega)$, 
as well as ${\rm Re}(n|\hat{U}_4|n)$.
There is approximate positive/negative symmetry except for the IR region 
in the distribution of the transverse matrix elements 
${\rm Im}(\omega^{-1}(n|\hat{U}_4|n)_\omega)$,
as well as ${\rm Im}(n|\hat{U}_4|n)$.
Here, the asymmetry in the IR region of 
${\rm Im}(\omega^{-1}(n|\hat{U}_4|n)_\omega)$ 
gives almost no influence on the Polyakov loop 
because of the damping factor $\lambda_n^{N_4-1}$. 
Thus, the matrix elements 
$\omega^{-1}(n|\hat{U}_4|n)_\omega$ and $(n|\hat{U}_4|n)$ 
have similar features, 
in spite of some difference in the shape of the distribution.
%
The results for the $Z_3$-rotated vacuum by $\omega^2$ are similar to 
those for the $\omega$-rotated one, 
which is natural because of the complex conjugation symmetry.

\section{Summary and concluding remarks}
In this study, we analytically and numerically have discussed 
the relation between confinement and chiral symmetry breaking 
based on the lattice QCD formalism. 
First, we derive the analytical relation (\ref{RelOrig}) connecting the Polyakov loop and Dirac modes 
on the temporally odd-number lattice with the normal periodic boundary condition for link-variables. 
Since the Polyakov loop is an order parameter of quark confinement and 
low-lying Dirac modes are essential for chiral symmetry breaking, 
this relation is useful for discussing the relation between confinement 
and chiral symmetry breaking. 
This relation is valid not only at the quenched level also but in the full QCD and in finite temperature/density. 
It is expected from the relation (\ref{RelOrig}) that 
low-lying Dirac modes have little contribution to the Polyakov loop. 

The numerical costs, in general, for solving the Dirac eigenequation are very large. 
On even lattice, where all the lattice sizes are even number, the numerical cost can be reduced 
by using the KS formalism. 
Although the KS formalism is not directly applicable 
to the temporally odd-number lattice, 
we have developed the modified KS formalism applicable to the temporally odd-number lattice. 
Using the modified KS formalism, we derive the relation (\ref{RelKS}) which is equivalent 
to the original relation (\ref{RelOrig}). 
Thus the numerical cost can be reduced on the temporally odd-number lattice. 

Next, we have performed the numerical lattice QCD Monte Carlo calculation 
with the standard plaquette action at the quenched level in both confinement and deconfinement phases. 
Of course, we impose the periodic boundary condition to the temporally odd-number lattice. 
Then we have numerically confirmed that the relation (\ref{RelOrig}) exactly holds and 
low-lying Dirac modes have little contribution to the Polyakov loop 
in both confinement and deconfinement phases, 
where the damping factor $\lambda_n^{N_4-1}$ in the relation (\ref{RelOrig}) plays an important role. 
These facts are observed similarly using the $Z_3$-rotated gauge configurations. 
Thus, we conclude that the relation between confinement and chiral symmetry breaking 
is not one-to-one correspondence in QCD. 

Also, we have investigated the property 
of the Dirac-mode matrix element $(n|\hat{U}_4|n)$ 
which appears in the relation (\ref{RelKS}). 
In the confinement phase, there is the positive/negative symmetry 
in the distribution of the matrix element $(n|\hat{U}_4|n)$, 
and hence the Polyakov loop is zero. 
In the deconfinement phase, however, 
the positive/negative symmetry disappears in the distribution of the matrix element $(n|\hat{U}_4|n)$, 
and then the Polyakov loop is nonzero. 
Corresponding to this, after the $Z_3$ rotation in the deconfinement phase, 
the distribution of the transverse matrix elements has the positive/negative symmetry 
while the distribution of the longitudinal matrix elements does not. 
However, the transverse matrix elements have asymmetry in the IR region of Dirac eigenvalues. 
Fortunately, this asymmetry does not affect the Polyakov loop 
because of the damping factor $\lambda_n^{N_4-1}$ in Eq. (\ref{RelKS}). 
In this way, we have discovered a new symmetry of the matrix element $(n|\hat{U}_4|n)$, 
which distinguishes confinement and deconfinement phases 
like the center symmetry in the pure-gauge theory. 
Thus, it is interesting to investigate the relation 
between the positive/negative symmetry and the center symmetry, 
which is very related to confinement \cite{Greensite}. 

In this study, we have performed the numerical analysis at the quenched level. 
However, the full QCD calculation is desired for more quantitative discussion. 
In particular, it is interesting to investigate the properties of the new positive/negative symmetry 
of the matrix element $(n|\hat{U}_4|n)$ in the full QCD calculation. 

Recently, the importance of the ratio of susceptibility of the Polyakov loop 
for the deconfinement transition was pointed out. 
Strictly speaking, the Polyakov loop must be renormalized 
for the physical continuum limit. 
However, one can discuss the deconfinement phase transition 
by considering the ratio of susceptibility of the Polyakov loop 
without uncertainties of renormalization of the Polyakov loop. 
We are now investigating the relation between 
confinement and chiral symmetry breaking 
using the ratio of the susceptibility of the Polyakov loop \cite{Redlich}. 

Also, it is interesting to study the relation between the QCD monopole and low-lying Dirac modes 
by using gauge-invariant Dirac-mode expansion \cite{GIS}. 
This is because the QCD monopole in the maximally Abelian gauge is important for 
non-perturbative phenomena of low-energy QCD, 
such as confinement and chiral symmetry breaking \cite{Miyamura, Woloshyn}. 

Finally, we note consequence of our conclusion of possible difference 
between confinement and chiral symmetry breaking in QCD, 
which our study indicates. 
These results imply that QCD can show a new phase, 
where chiral symmetry is restored but the quark is confined 
\cite{YAoki, LS11, GIS}. 
For example, non-trivial effects of strong electromagnetic fields on chiral symmetry 
can change the structure of the QCD vacuum \cite{SuganumaTatsumi}. 

\section*{Acknowledgements}
The authors thank Professor K. Redlich and Dr. C. Sasaki for valuable discussions and comments. 
T.M.D. thanks H. Iida, N. Yamanaka and S. Imai for useful discussions and comments. 
H. S. is supported in part by the Grant for Scientific Research [(C) No.23540306, E01:21105006] 
from the Ministry of Education, Science and Technology of Japan. 
The lattice QCD calculations were performed on the NEC-SX8R 
and NEC-SX9 at Osaka University.

\appendix
\section{\label{EvenDerivation}
Derivation of a relation between the Polyakov loop and Dirac modes on the even lattice}
In this paper, we consider the temporally odd-number lattice, 
derive the analytical relation connecting the Polyakov loop and Dirac modes 
and discuss the relation between confinement and chiral symmetry breaking. 
In this section, however, we derive the relation between the Polyakov loop and Dirac modes 
on the even lattice where all the lattice sizes are even number 
with the periodic boundary condition for link-variables. 

\begin{figure}[h]
\begin{center}
\includegraphics[scale=0.4]{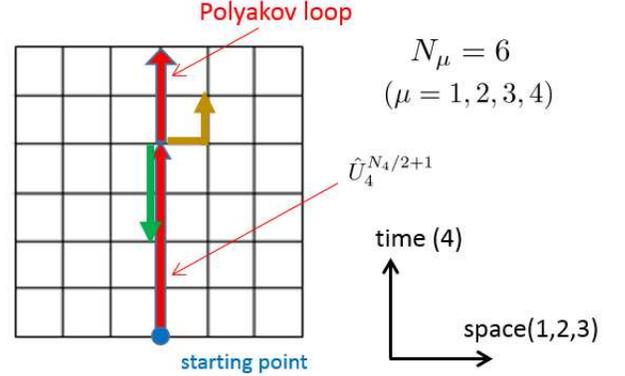}
\caption{
An example of even lattice. 
This is the $N_\mu=6\ (\mu=1,2,3,4)$ case. 
Each line corresponds to each term in $\hat{U}_4^{N_4/2+1}\hat{\slashb{D}}^{N_4/2-1}$ in Eq.(\ref{Itilde}). 
}
\label{EvenLattice}
\end{center}
\end{figure}

First, corresponding to $I$ in Eq.(\ref{I}), we introduce 
\begin{align}
\tilde{I}(N_4)\equiv{\rm Tr}_{c,\gamma}(\gamma_4^{\xi(N_4)}\hat{U}_4^{N_4/2+1}\hat{\slashb{D}}^{N_4/2-1}), 
\label{Itilde} 
\end{align}
where $\xi(N_4)$ is defined as 
\begin{align}
\xi(N_4)=\begin{cases}
0 & (N_4/2:{\rm odd}) \\
1 & (N_4/2:{\rm even})
\end{cases}
\label{xi}
\end{align}
Like the case of the temporally odd-number lattice, 
$\hat{U}_4^{N_4/2+1}\hat{\slashb{D}}^{N_4/2-1}$ is expressed as 
a sum of products of $N_4$ link-variable operators. 
In Fig. \ref{EvenLattice}, an example of the even lattice is shown and 
each line corresponds with each term in $\hat{U}_4^{N_4/2+1}\hat{\slashb{D}}^{N_4/2-1}$ 
in Eq.(\ref{Itilde}). 
Note that there are no closed loops in $\hat{U}_4^{N_4/2+1}\hat{\slashb{D}}^{N_4/2-1}$ 
because the number of $\hat{U}_4$ is larger than that of $\hat{U}_{-4}$. 
Thus, 
$\hat{U}_4^{N_4/2+1}\hat{\slashb{D}}^{N_4/2-1}$ does not have any operators corresponding to closed paths 
except for the term proportional to $\hat{U}_4^{N_4}$, which is proportional to the Polyakov loop. 
Therefore using the periodic boundary condition for temporal direction and 
Eqs.(\ref{PolyakovOp}) and (\ref{nonclosed}), 
we obtain 
\begin{align}
\tilde{I}
&={\rm Tr}_{c,\gamma} (\gamma_4^{\xi(N_4)}\hat{U}_4^{N_4/2+1}\hat{\slashb{D}}^{N_4/2-1}) \nonumber \\
&={\rm Tr}_{c,\gamma} \{\gamma_4^{\xi(N_4)}\hat{U}_4^{N_4/2+1} (\gamma_4 \hat{D}_4)^{N_4/2-1}\} \nonumber \\
&={\rm Tr}_{c,\gamma} (\gamma_4^{\xi(N_4)+N_4/2-1}\hat{U}_4^{N_4/2+1} \hat{D}_4^{N_4/2-1}) \nonumber \\
&=4 {\rm Tr}_{c} (\hat{U}_4^{N_4/2+1} \hat{D}_4^{N_4/2-1}) \nonumber \\
&=\frac{4}{(2a)^{N_4/2-1}}{\rm Tr}_{c} \{\hat{U}_4^{N_4/2+1} (\hat{U}_4-\hat{U}_{-4})^{N_4/2-1}\} \nonumber \\
&=\frac{4}{(2a)^{N_4/2-1}}{\rm Tr}_{c} \{ \hat{U}_4^{N_4} \} \nonumber \\
&=\frac{12V}{(2a)^{N_4/2-1}}L_P. \label{Itilde1}
\end{align}

On the other hand, 
taking Dirac modes as the basis for the functional trace in Eq.(\ref{Itilde}), 
we find 
\begin{align}
\tilde{I}
&=\sum_n\langle n|\gamma_4^{\xi(N_4)}\hat{U}_4^{N_4/2+1}\slashb{\hat{D}}^{N_4/2-1}|n\rangle \nonumber\\
&=i^{N_4/2-1}\sum_n\lambda_n^{N_4/2-1}
\langle n|\gamma_4^{\xi(N_4)}\hat{U}_4^{N_4/2+1}| n \rangle.  \label{Itilde2}
\end{align}

Combining Eqs.(\ref{Itilde1}) and (\ref{Itilde2}), we derive the relation between 
the Polyakov loop $L_P$ 
and the Dirac eigenvalues $i\lambda_n$ on the even lattice: 
\begin{eqnarray}
L_P
=\frac{(2ai)^{N_4/2-1}}{12V}
\sum_n\lambda_n^{N_4/2-1}\langle n|\gamma_4^{\xi(N_4)}\hat{U}_4^{N_4/2+1}| n \rangle. \ \ \ \ 
\label{RelOrigEven}
\end{eqnarray}

Comparing Eqs.(\ref{RelOrig}) and (\ref{RelOrigEven}), 
Eq.(\ref{RelOrig}) is more simple than Eq.(\ref{RelOrigEven}). 
However, physics should not depend on temporal lattice size $N_4$ and 
damping factor $\lambda_n^{N_4/2-1}$ is expected to have essential role in RHS of Eq.(\ref{RelOrigEven}) 
like the case of the temporally odd-number lattice. 

\section{\label{CalMatEle}
The relation between the Dirac matrix element and the KS Dirac matrix element}
In this section, 
we consider the relation between the Dirac matrix element and the KS Dirac matrix element 
in both even and temporally odd-number lattices. 

\subsection{\label{CalMatEleEven} The case of the even lattice}
First, we consider the even lattice and the original KS formalism. 
Using Eq.(\ref{PsiChiEven2}), 
the Dirac matrix element of a link variable operator $\langle n,I|\hat{U}_\mu| m,J \rangle$ 
can be expressed by the KS Dirac eigenfunction as 
\begin{align}
&\langle n,I|\hat{U}_\mu| m,J \rangle \nonumber \\
&=\sum_{s,\alpha}\psi_n^I(s)^\dagger_\alpha U_\mu(s) \psi_m^J(s+\hat{\mu})_\alpha \nonumber \\
&=\sum_{s,\alpha}\chi_n(s)^\dagger T^\dagger(s)_{I\alpha} U_\mu(s) T(s+\hat{\mu})_{\alpha J}\chi_m(s+\hat{\mu})
 \nonumber \\
&=\sum_{s}\chi_n(s)^\dagger \{T^\dagger(s)T(s+\hat{\mu})\}_{IJ}U_\mu(s)\chi_m(s+\hat{\mu}).  \label{UmuDiracKSEven1}
\end{align}
$T^\dagger(s)T(s+\hat{\mu})$ can be calculated from the definition of the matrix $T(s)$ (\ref{T}): 
\begin{align}
T^\dagger(s)T(s+\hat{\mu})
=
\tilde{\eta}^{({\rm E})}_\mu(s)\gamma_\mu, \label{TT}
\end{align}
where $\tilde{\eta}^{({\rm E})}_\mu(s)$ is a sign function defined as 
\begin{align}
\tilde{\eta}^{({\rm E})}_\mu(s)=(-1)^{s_{\mu+1}+\cdots+s_4} \ (\mu\leq3), \ \ \ \ \ 
\tilde{\eta}^{({\rm E})}_4(s)=1, \label{ModStagEven}
\end{align}
which is similar to the staggered phase (\ref{eta}). 
Thus, the Dirac matrix element is expressed as 
\begin{align}
&\langle n,I|\hat{U}_\mu| m,J \rangle \nonumber\\
&=(\gamma_\mu)_{IJ}
\sum_{s}\tilde{\eta}^{({\rm E})}_\mu(s)\chi_n(s)^\dagger U_\mu(s) \chi_m(s+\hat{\mu}) \nonumber\\
&=(\gamma_\mu)_{IJ}
(n|\hat{\tilde{\eta}}^{({\rm E})}_\mu \hat{U}_\mu|m), 
\label{UmuDiracKSEven2}
\end{align}
where $\hat{\tilde{\eta}}^{({\rm E})}_\mu$ is an operator defined as 
\begin{align}
\langle s|\hat{\tilde{\eta}}^{({\rm E})}_\mu |s'\rangle=\tilde{\eta}^{({\rm E})}_\mu(s)\delta_{ss'}
\end{align}
corresponding to the sign function $\tilde{\eta}^{({\rm E})}_\mu(s)$. 
In particular, since $\tilde{\eta}^{({\rm E})}_4(s)=1$ is satisfied for $\mu=4$, we obtain 
\begin{align}
\langle n,I|\hat{U}_4| m,J \rangle
=(\gamma_4)_{IJ}(n|\hat{U}_4| m). \label{U4DiracKSEven}
\end{align}

Not only the Dirac matrix element of a single link-variable operator, 
$\langle n,I|\hat{U}_\mu| m,J \rangle$, 
but also that of any operator consisting of link-variable operators, 
$\langle n,I|\hat{O}(\hat{U})| m,J \rangle$, can be evaluated in terms of 
the KS Dirac matrix element or the KS Dirac eigenfunction $\chi_n(s)$ 
by a similar calculation on the even lattice. 

\subsection{\label{CalMatEleOdd} The case of the temporally odd-number lattice}
Next, we consider the temporally odd-number lattice and the modified KS formalism. 
Like the case of the even lattice, using Eq.(\ref{PsiChiOdd2}), 
the Dirac matrix element of a link variable operator $\langle n,I|\hat{U}_\mu| m,J \rangle$ 
can be expressed by the KS Dirac eigenfunction as 
\begin{align}
&\langle n,I|\hat{U}_\mu| m,J \rangle \nonumber\\
&=\sum_{s,\alpha}\psi_n^I(s)^\dagger_\alpha U_\mu(s) \psi_m^J(s+\hat{\mu})_\alpha \nonumber\\
&=\sum_{s,\alpha}\chi_n(s)^\dagger M^\dagger(s)_{I\alpha} U_\mu(s) M(s+\hat{\mu})_{\alpha J}\chi_m(s+\hat{\mu}) \nonumber\\
&=\sum_{s}\chi_n(s)^\dagger \{M^\dagger(s)M(s+\hat{\mu})\}_{IJ}U_\mu(s)\chi_m(s+\hat{\mu}) 
\label{UmuDiracKSOdd1}
\end{align}
Corresponding to Eq.(\ref{TT}), 
$M^\dagger(s)M(s+\hat{\mu})$ is expressed as 
\begin{align}
M^\dagger(s)M(s+\hat{\mu})
=
\tilde{\eta}^{({\rm O})}_\mu(s)\gamma_\mu \gamma_4, 
\end{align}
where $\tilde{\eta}^{({\rm O})}_\mu(s)$ is a sign function defined as 
\begin{align}
\tilde{\eta}^{({\rm O})}_\mu(s)=(-1)^{s_1+\cdots+s_\mu} \ (\mu\leq3), \ \ \ \ \ 
\tilde{\eta}^{({\rm O})}_4(s)=1, \label{ModStagOdd}
\end{align}
which is different from both the staggered phase $\eta_\mu(s)$ 
and the sign function in the even lattice $\tilde{\eta}^{({\rm E})}_\mu(s)$. 
Thus, the Dirac matrix element is expressed as 
\begin{align}
&\langle n,I|\hat{U}_\mu| m,J \rangle \nonumber\\
&=(\gamma_\mu \gamma_4)_{IJ}\sum_{s}\tilde{\eta}^{({\rm O})}_\mu(s)\chi_n(s)^\dagger U_\mu(s) \chi_m(s+\hat{\mu}) \nonumber\\ 
&=(\gamma_\mu \gamma_4)_{IJ}(n|\hat{\tilde{\eta}}^{({\rm O})}_\mu \hat{U}_\mu|m),
\label{UmuDiracKSOdd2}
\end{align}
where $\hat{\tilde{\eta}}^{({\rm O})}_\mu$ is an operator defined as 
\begin{align}
\langle s|\hat{\tilde{\eta}}^{({\rm E})}_\mu |s'\rangle=\tilde{\eta}^{({\rm E})}_\mu(s)\delta_{ss'}
\end{align}
corresponding to the sign function $\tilde{\eta}^{({\rm O})}_\mu(s)$. 
In particular, since $\tilde{\eta}^{({\rm O})}_4(s)=1$ is satisfied for $\mu=4$, we obtain 
\begin{align}
\langle n,I|\hat{U}_4| m,J \rangle
=\delta_{IJ}(n|\hat{U}_4| m). \label{U4DiracKSOdd}
\end{align}
For more special case, the diagonal component $\langle n,I|\hat{U}_4| n,I \rangle$ is expressed as 
\begin{align}
\langle n,I|\hat{U}_4| n,I \rangle
=(n|\hat{U}_4| n). \label{U4DiracKSOddDiagApp}
\end{align}

Like the case of the even lattice, one can evaluate the Dirac matrix element of the other operator 
$\langle n,I|\hat{O}(\hat{U})| m,J \rangle$ 
using the KS Dirac matrix element or the KS Dirac eigenfunction $\chi_n(s)$ 
on the temporally odd-number lattice.

\end{document}